\documentclass[12pt]{oxarticle}
\pdfoutput=1

\usepackage{graphicx, float, array, xspace, amscd, amsmath, amsthm, amssymb, latexsym, relsize, bbm}
\usepackage[a4paper, left=1.7in, right=0.3in, top=0.8in, bottom=1.2in]{geometry}
\usepackage[bbgreekl]{mathbbol}
\usepackage{amsfonts, tikz-cd}
\usepackage[T1]{fontenc}
\PassOptionsToPackage{linecolor=blue,backgroundcolor=blue!25,bordercolor=blue,textsize=scriptsize}{todonotes}
\usepackage{todonotes}
\usepackage{tikz}
\usetikzlibrary{backgrounds}
\usepackage[framemethod=tikz]{mdframed}
\AtBeginEnvironment{mdframed}{%
\tikzset{every picture/.style={}}%
}
\mdfsetup{roundcorner=.5ex}
{\begin{mdframed}[backgroundcolor=white]}%
{\end{mdframed}}

\DeclareSymbolFontAlphabet{\mathbb}{AMSb}
\DeclareSymbolFontAlphabet{\mathbbl}{bbold}

\usepackage[sort&compress, comma, square, numbers]{natbib}
\usepackage[all,cmtip]{xy}
\usepackage{color}
\definecolor{MyDarkBlue}{rgb}{0.15,0.25,0.45}
\usepackage[linktocpage=true]{hyperref}
\hypersetup{colorlinks=true, pageanchor=false, citecolor=blue, linkcolor=blue, urlcolor=blue, pdfauthor={}, pdftitle={The heterotic superpotential and the composite Hull connection},breaklinks=true}

\let\SS=\S 

\def\Dbar{{\overline{D}}}

\newcommand{\ol}[1]{{\overline{#1}}}

\newcommand{\LC}{\text{\tiny LC}}

\newcommand{\Ch}{\text{\tiny CH}}

\renewcommand{\sb}{{\overline{\sigma}}}

\newcommand{\kb}{{\overline{ \kappa}}}
\newcommand{\rb}{{\overline{ r}}}

\newcommand{\w}{{\,\wedge\,}}

\newcommand{\wh}{\widehat}

\newcommand{\half}{\frac{1}{2}}

\def\CS{{\text{CS}}}

\renewcommand{\a}{\alpha}

\newcommand{\G}{\Gamma}
\renewcommand{\d}{\delta}\newcommand{\D}{\Delta}

\newcommand{\Th}{\Theta}

\newcommand{\m}{\mu}
\newcommand{\n}{\nu}

\renewcommand{\r}{\rho}
\newcommand{\s}{\sigma}\renewcommand{\S}{\Sigma}

\renewcommand{\o}{\omega}\renewcommand{\O}{\Omega}

\DeclareFontFamily{OT1}{pzc}{}
\DeclareFontShape{OT1}{pzc}{m}{it}{<-> s * [1.200] pzcmi7t}{}
\DeclareMathAlphabet{\mathpzc}{OT1}{pzc}{m}{it}

\newcommand{\cB}{\mathcal{B}}\newcommand{\ccB}{\mathpzc B}

\newcommand{\cE}{\mathcal{E}}

\newcommand{\cK}{\mathcal{K}}

\newcommand{\cO}{\mathcal{O}}

\newcommand{\cS}{\mathcal{S}}

\newcommand{\ccZ}{\mathpzc Z}

\DeclareFontFamily{U}{bbold}{}
\DeclareFontShape{U}{bbold}{m}{n}
 {  <-5.5> s*[1.05] bbold5
    <5.5-6.5> s*[1.05] bbold6
    <6.5-7.5> s*[1.05] bbold7
    <7.5-8.5> s*[1.05] bbold8
    <8.5-9.5> s*[1.05] bbold9
    <9.5-11.5> s*[1.05] bbold10
    <11.5-16> s*[1.05] bbold12
    <16-> s*[1.05] bbold17
 }{}

\newcommand{\IR}{\mathbbl{R}}

\font\eightrm=cmr8 at 8pt
\font\csc=cmcsc10

\newcommand{\beq}{\begin{equation}}
\newcommand{\eeq}{\end{equation}}
\newcommand{\beqnn}{\begin{equation*}}
\newcommand{\eeqnn}{\end{equation*}}
\newcommand{\bea}{\begin{eqnarray}}
\newcommand{\eea}{\end{eqnarray}}
\newcommand{\bean}{\begin{eqnarray*}}
\newcommand{\eean}{\end{eqnarray*}}

\newcommand{\sref}[1]{\SS\ref{#1}}

\newcommand{\ii}{\text{i}}

\newcommand{\place}[3]{\vbox to0pt{\kern-\parskip\kern-7pt
                             \kern-#2truein\hbox{\kern#1truein #3}
                             \vss}\nointerlineskip}

\DeclareFontFamily{U}{wncy}{}
\DeclareFontShape{U}{wncy}{m}{n}{<->wncyr10}{}
\DeclareSymbolFont{mcy}{U}{wncy}{m}{n}
\DeclareMathSymbol{\sha}{\mathord}{mcy}{"58}

\newcommand{\del}{{\partial}}
\newcommand{\delb}{{\overline{\partial}}}

\newcommand{\lb}{{\overline\lambda}}
\newcommand{\nb}{{\overline\n}}
\newcommand{\mb}{{\overline\m}}

\renewcommand{\aa}{\mathfrak{a}}

\newcommand{\dd}{{\text{d}}}

\newcommand{\K}{K\"ahler\xspace}

\renewcommand{\H}{\text{H}}

\newcommand{\tr}{\text{tr}\hskip2pt}

\newcommand{\tb}{{\overline{\tau}}}

\newcommand{\ap}{{\a^{\prime}\,}}
\renewcommand{\sb}{{\overline{\sigma}}}

\renewcommand{\rb}{{\overline{\rho}}}
\renewcommand{\=}{\;=\;}

\makeatletter
\g@addto@macro\bfseries{\boldmath}
\makeatother

\newcommand{\citeUG}{\cite{Candelas:2018lib, McOrist:2019mxh}\xspace}

\newcommand{\citeAQS}{\cite{Anguelova:2010ed}\xspace}

\newcommand{\citeMMS}{\cite{Melnikov:2014ywa}\xspace}

\newcommand{\citeUGAP}{\cite{McOrist:2026boe}\xspace}

\newcommand{\citeBdRtwo}{\cite{Bergshoeff:1988nn, Bergshoeff:1989de}\xspace}
\newcommand{\citeBdR}{\cite{ Bergshoeff:1989de}\xspace}

\newcommand{\citeMP}{\cite{McOrist:2025zwf}\xspace}

\newcommand{\citeapthree}{\cite{MPap3}\xspace}

\renewcommand{\baselinestretch}{1.1}
\numberwithin{equation}{section}
\proofmodefalse

\begin{document}
\pagestyle{empty}
\begin{center}
\null\vskip0.2in {\Huge  The heterotic superpotential and the $\alpha'$ expansion \\[0.5in]}

{\csc Jock McOrist$^{1}$ \\[0.2in]}

{\it Department of Mathematics\hphantom{$^2$}\\ School of Science and Technology\\ University of New England\\ Armidale, 2351, Australia\\[3ex]   }

\footnotetext[1]{{\tt  jmcorist@une.edu.au} }

\vspace{0.4cm}

\vfill
{\bf Abstract\\[-12pt]}
\end{center} 

We study the role of the Hull connection in the heterotic superpotential  for $N=1$ $d=4$ vacua. The integrand is defined in terms of $H$, which is in turn defined in terms of a gravitational Chern--Simons coupling involving $H$. We show this recursive structure is key to the superpotential reproducing some of the ten--dimensional supersymmetry conditions written down by Bergshoeff--de Roo through $\ap^3$. With this evidence in hand we make some observations about extending some of the tensor equations of Hull and Strominger to all orders in $\alpha'$.

\vskip150pt

\vfill
\newgeometry{left=1.5in, right=0.5in, top=0.75in, bottom=0.8in}
\setcounter{page}{1}
\hypersetup{pageanchor=true}
\pagestyle{plain}
\renewcommand{\baselinestretch}{1.3}\selectfont
\newpage

\section{Introduction}
\label{sec:introduction}

Consider heterotic string theory on $\IR^{3,1}\times X$ with an $E_8\times E_8$ gauge bundle preserving $N = 1$ $d = 4$ supersymmetry. The microscopic approach to formulating its supersymmetry conditions is to dimensionally reduce the ten-dimensional fermionic variations in \citeBdRtwo to four-dimensions.  This was first done by \cite{Hull:1986kz,Strominger:1986uh} to first order in $\ap$ and resulted in a set of tensor equations on an  $SU(3)$ structure on $X$, which includes
\beq\label{eq:leadingF}
\dd \O \= 0~, \qquad H \= \ii (\del - \delb)\o~, \qquad F^{0,2} \= 0~,
\eeq
where $X$ is complex with $\omega$ the hermitian form on $X$ and $\O$ the compatible holomorphic volume form. These are supplemented by a heterotic Bianchi identity
\beq\label{eq:BianchiH}
\dd H \= \frac{\ap}{4} \tr F\w F - \frac{\ap}{4} \tr R^\H \w R^\H~,
\eeq
where $R^\H$ is the curvature for a non-hermitian connection on the tangent bundle and $F$ is the curvature for the connection on the gauge bundle. There are two additional equations: the metric is conformally balanced and $F$  satisfies a hermitian Yang--Mills equation. 

The  heterotic superpotential \cite{Becker:2003gq,LopesCardoso:2003dvb,Gurrieri:2004dt} 
\beq
\label{eq:superpotential}
W \= \int_X(H + \ii\dd\o)\w\Omega~.
\eeq
 is a functional whose job is to reproduce a subset of the underlying ten--dimensional supersymmetry conditions. In the classical limit, in which the composite nature of the tangent bundle connection is suppressed, the  stationary equations $\d W = 0$ and $W=0$ reproduce \eqref{eq:leadingF}. This is why \eqref{eq:leadingF}  are often referred to as as `F-terms',  in analogy with $d = 4$ supersymmetric quantum field theory. The conformally balanced and Hermitian Yang--Mills equation for $F$ are often referred to as D-terms for reasons we discuss further below.
 
Beyond the classical limit, the advantage of heterotic string theory is that we have available the ten-dimensional supersymmetry analysis of Bergshoeff--de Roo (BdR) \citeBdRtwo. These  results completely fix the $\ap^2$ corrections to the action and supersymmetry transformations and partially fix the $\ap^3$ corrections.\footnote{At $\ap^3$, the BdR construction completely fixes the corrections required by supersymmetrising the Yang--Mills and Lorentz Chern--Simons terms. It misses the gravitational multiplet that starts with $\ap^3\zeta(3)R^4$ term and its supersymmetric completion, see also \cite{deRoo:1992zp}.} In \citeMP these transformations were applied to an SU(3) manifold, and in their original field basis the equations \eqref{eq:leadingF} receive $\ap^2$ corrections and $\ap^3$ corrections \citeapthree. At first sight, this seems at odds with the expectation that the superpotential receives no perturbative $\ap$ corrections. Our goal here is to ask a simple question: is the usual expression for $W$ compatible with these $\ap$--corrected supersymmetry equations?

There is a  point to keep in mind. Four-dimensional $N \= 1$ perturbative non-renormalisation statements concern a superpotential written in chiral fields. The relation between the $d=4$ chiral superfields and the ten-dimensional fields can itself receive $\ap$ corrections. Perturbative non-renormalisation  does not obviously apply to the expression for $W$ in the six-dimensional fields. Our approach here is to take $W$ in \eqref{eq:superpotential} as a candidate functional and test its stationary equations against the known supersymmetry conditions. This will tell us about the non-renormalisation properties of $W$ as a six dimensional integral. Keeping   \eqref{eq:superpotential} unchanged  does not necessarily imply its stationary locus is unchanged. The dependence of $H$ on the  Hull connection generates  $\ap$ corrections when we vary the underlying fields. 

The main results of this paper are as follows. We first compute the stationary loci of \eqref{eq:superpotential} in the original BdR field basis \citeBdRtwo. Key to this is that $H$ is defined in terms of the Chern--Simons form fo the composite Hull connection and so gives a recursion relation. We solve this recursion relation by rewriting expressions in terms of a geometric series in $\ap$, which can be summed\footnote{The recursive role of $H$ has been discussed in the context of generalised geometry in \cite{Coimbra:2014geometry}. It would be interesting to connect the work there with our approach. } This allows us to write down the stationary points of $W$ in closed form.  We compare the equations at $\ap^2$ and $\ap^3$ and find an exact match to the supersymmetry equations of BdR \citeBdR. This forms a non-trivial check that $W$ and BdR are consistent. 

In \citeMP, it was shown that many of these $\ap^2$ corrections can be neatly absorbed by a field redefinition of the hermitian form and dilaton. What we did not know a priori was whether these corrections would follow from $W$. Our calculation shows that they do: the stationary equations of $W$ match the corresponding ten-dimensional supersymmetry equations in the original fields.  There is an interesting comparison with type II string theory. It is conjectured in \cite{Smith:2024moduli} that the analogous off-shell bulk superpotential receives no perturbative $\ap$ corrections. The bulk corrections in type II start at $\ap^3$, but their full supersymmetric completion for general flux backgrounds is not known, see for example \cite{liu2025typeiibderivativesfivepoint}, making a comparable direct check more difficult. A related check was made in \cite{Ashmore:2026vgf}, where an off-shell real superpotential reproduces the eleven-dimensional supersymmetry conditions for M-theory on eight-manifolds at the two-derivative level.   Our heterotic calculation is consistent with this non-renormalisation picture of $W$: through $\ap^3$ in the Chern--Simons sector, the usual expression for $W$ already produces the corrections required by ten-dimensional supersymmetry. 

 While $W$ is not renormalised, we do find that the stationary points or `F-terms' do receive $\ap$ corrections. 
For example, the second equation in \eqref{eq:leadingF}, $H=i(\del-\delb)\omega$, is not by itself a stationary equation of $W$, already at $\ap^2$. This becomes even more  more pronounced at $\ap^3$.  But we do show that they agree with the results in \citeMP and \citeapthree after imposing the D-terms. We also compare the stationary equations of $W$ at $\ap^3$ with the BdR spinor transformations, using results from upcoming work \citeapthree. We again find complete agreement up to D-terms. A related comment is that the holomorphic volume form $\Omega$ is renormalised.   We obtain an expression for a fully $\ap$--corrected three-form $\wh\Omega$ and show that $\dd\wh\Omega=\cO(\ap^4)$ on the stationary locus. It matches precisely the volume form in \citeMP when truncated to $\ap^2$. With a mild assumption about holomorphy of $W$, we furthermore show $\dd \wh \O = 0$ to all orders in $\ap$.

This all gives good evidence that $W$ is to be taken seriously as a functional for determining supersymmetry conditions. We then use it to write down some equations, correct to all orders in $\ap$. These form conjectures for how some of the supersymmetry equations of \cite{Bergshoeff:1989de,McOrist:2025zwf,MPap3} are extended to all orders in $\ap$. For example, we find the very sensible equation that $F^{0,2}_{\wh J} = 0$ where $\wh J$ is the complex structure determined by $\wh \Omega$.

Flux superpotentials written as integrals over the internal manifold were introduced in F/M--theory in \cite{Gukov:1999ya}. In heterotic theory, the flux superpotential was proposed by studying dimensional reduction on a Calabi--Yau threefold in \cite{Becker:2002jj} and studied in non-\K compactifications in \cite{Becker:2003yv}. Its relation to a rewriting of the ten-dimensional action in terms of BPS conditions was developed in \cite{LopesCardoso:2003dvb}, while the general $SU(3)$-structure expression was derived at leading order in $\ap$ from the gravitino mass term in \cite{Gurrieri:2004dt}. Its normalisation was fixed by matching matter Yukawa couplings obtained by dimensional reduction in \cite{McOrist:2016cfl}. In previous studies of $W$, the composite nature of the connection has not been addressed; this is our primary goal in this paper. The superpotential \eqref{eq:superpotential} and its possible non-perturbative corrections are also discussed in \cite{Sethi:2017phn}. 

The relation of the heterotic superpotential to infinitesimal moduli was developed in \cite{delaOssa:2015maa} and extended to finite deformations and their $L_\infty$ structure in \cite{Ashmore:2018ybe}. A generalised-geometry formulation of the superpotential and \K potential was given in \cite{Ashmore:2019rkx}, with supersymmetry encoded by involutivity and the vanishing of a moment map. The relation of the superpotential and BPS complexes were recently discussed in \cite{Kupka:2024rvl} and have been used to study analogues of Kodaira--Spencer theory, see for example \cite{Ashmore:2023vji}. These analyses use the same form of $W$, but treat the tangent-bundle connection entering $H$ as an  independent auxiliary field. Here we keep the dependence of the Hull connection on the metric and $H$ when varying $W$. This allows us to match the corresponding $\ap^2$ and $\ap^3$ corrections of \citeBdR, up to lower-order D-terms.  The superpotential also gives a $\Dbar$-operator describing infinitesimal deformations \cite{delaOssa:2015maa}, reproducing the earlier direct construction in \cite{delaOssa:2014cia}. Both descriptions include auxiliary tangent-bundle deformations. These were eliminated to first order in $\ap$ in \cite{McOrist:2021dnd} and using the moduli space metric developed in \cite{Anguelova:2010ed, Candelas:2016usb, McOrist:2019mxh} an adjoint $\Dbar^\dag$ was constructed whose kernel being consistent with the conformally balanced and HYM equations being imposed. This is one reason why we call these two equations `D-terms'. This description was developed further in \cite{McOrist:2024zdz}, where the supersymmetry equations and Bianchi identity were related to a Hermitian Yang--Mills condition on the extension bundle, again to first order in $\ap$. It would be interesting to use the results here to construct a $\Dbar$ operator valid to all orders in $\ap$, but we leave this for future work.

In \sref{sec:superpotential}, we study variations of the superpotential with the Hull connection.  We then check the $\ap^2$ in \sref{sec:BdR-checks} and $\ap^3$  equations in \sref{sec:cubic} against BdR  and find agreement. In \sref{s:holomorphy}, we discuss holomorphy of the superpotential. In \sref{sec:all-orders} we give some all order in $\ap$ F-term results. The appendices give the component calculations and a short comparison with four-dimensional supersymmetry.

\section{The superpotential and its composite variation}
\label{sec:superpotential}
\label{s:superpotentialBdR}
\subsection{Fields and conventions}

Let $X$ be a compact complex threefold with Hermitian form $\o$ and a smooth nowhere-zero $(3,0)$-form $\Omega$. The fields have smooth expansions in $\ap$. We keep track of the $\ap$ dependence of both the equations and the background fields, as emphasised by \citeMMS.

For the comparison with supersymmetry, $\Omega$ is the rescaling by the dilaton of the unit normalised spinor bilinear. In the normalisation of \citeMP,
\beq
\label{eq:Omega-spinor-definition}
\Omega \= e^{-2\Phi}\Psi~, \qquad \Psi_{mnp} \= \bar\eta^{T}\gamma_{mnp}\bar\eta~, \qquad \|\Psi\|_g \= 1~.
\eeq
Here $\eta$ is the unit internal spinor. We use $\Omega$ as the independent three-form when varying $W$, and use \eqref{eq:Omega-spinor-definition} to compare its stationary equations with supersymmetry.  

We work with the candidate superpotential \eqref{eq:superpotential}.
The Hull connection is the composite connection used by BdR \cite{Bergshoeff:1989de,Bergshoeff:1988nn}. We use $\G^\H \= \G^\LC + \half H$, $\G^- \= \G^\LC - \half H$ and $R_{mnpq} \= g_{pr}R_{mn}{}^r{}_q$. The curvature and Bianchi identities are
\beq
\label{eq:pair}
R^\H_{mnpq} \= R^-_{pqmn} + \half(\dd H)_{mnpq}~, \qquad \dd H \= \frac{\ap}{4}\tr(F\w F - R^\H\w R^\H)~.
\eeq
The role of the torsional connection in solving the heterotic Bianchi identity perturbatively was studied by Becker and Sethi \cite{Becker:2009df}.  

Real coordinate indices are $m,n,p,q$, and Greek indices are complex. We put $\dd^c \= \ii(\del - \delb)$. All complex types refer to $J$ unless stated otherwise. The trace is the matrix trace, so $\tr(\d\Th\w R) \= -\d\Th_{pq}\w R^{pq}$.

The variation of $H$ is
\beq
\label{eq:h-deformation}
\d H \=  \dd\cB + \frac{\ap}{2}\tr(\d A \w F) - \frac{\ap}{2}\tr(\d \Theta^\H\w R^\H) ~.
\eeq
Here $\cB$ includes deformations of the gauge field so that the total combination is invariant. Intuitively we think of $\ccB$ as the extension of $\d B$ to the heterotic setting. We combine this with the metric
\beq
\label{eq:conventions}
M_{mp} \= \half(\d g_{mp} + \cB_{mp})~.
\eeq
We choose a Lorentz frame in which the variation of the vielbein is $\d e_{ap} \= E_a{}^qM_{qp}$. In that frame, a variation of the Hull connection is 
\beq
\label{eq:variation}
\begin{aligned}
\d\Th^\H_{mpq}& \= \nabla^-_qM_{mp} - \nabla^-_pM_{mq} - H_{pq}{}^rM_{mr} + \frac{\ap}{4} \left( \tr(\d A\w F - \d\Th^\H\w R^\H) \right)_{mpq}~.\\[4pt]
\end{aligned}
\eeq

\subsection{Holomorphic fluctuations}
We consider holomorphic fluctuations of fields. The complexified hermitian form is written as 
$$
\ccZ \= \cB + \ii\d\o~, 
$$
while a holomorphic variation of the gauge connection is $\aa \= (\d A)^{0,1}$. The three-form varies as  $\d\Omega \= k\Omega + \D^\r\w\Omega_{\r}$, where $\Omega_{\r} \= \half\Omega_{\r\m\n}\dd x^\m\w\dd x^\n$ and $ k$ corresponds to rescaling $\O$.

We assume the background is complex but consider variations that take us away from this complex background. For example, although $\dd\o\w\Omega = 0$ on the background, we do not require its variation continue to vanish. Varying $W$, with  integration by parts gives
\beq
\label{eq:superpotential-variation}
\begin{aligned}
\d W &\=\!-\int_X\ccZ\w\dd\Omega + \frac{\ap}{2}\int_X \tr\big(\d A\w F - \d\Th^\H\w R^\H\big)\w\Omega + \int_X(H + \ii\dd\o)\w\d\Omega~. 
\end{aligned}
\eeq

At zeroth order in $\ap$, the $\ccZ^{0,2}$, $k$ and $\D$ coefficients give $\delb\Omega \= 0$, $H^{0,3} \= 0$ and $H^{1,2} \= -\ii\delb\o$. Hence, to zeroth order:\footnote{If we were to treat $W$ as a classical theory of gravity coupled to Yang-Mills in the spirit of \cite{Chapline:1982ww} without the Green-Schwarz anomaly cancellation, then $\ap$ is replaced by $g_{YM}^{-2}$ and classical stationary equation for $\aa$ gives $F^{0,2} = 0$. }
\beq\label{eq:fluxtorsionzero}
\dd \O \= 0~, \qquad H \= \dd^c\o + \cO(\ap)~.
\eeq

\subsection{Stationary points of $W$}
The key to this paper is in the recursive nature of $H$.  As $H$ is defined in terms of $\CS(\Th^\H)$ its variation is non-linear, equivalently recursive in nature. To illustrate this, consider just the gauge variation $\aa$ with all other fields fixed:
$$
\d_\aa W \= \int_X \d_\aa H \w \Omega~.
$$
 Since $\d_\aa\Th^\H_{mpq} \= \half(\d_\aa H)_{mpq}$, we have
\beq
\label{eq:W-gauge-recursion}
\d_\aa H \= \frac{\ap}{2}\tr(\aa\w F) + \frac{\ap}{4}R^{\H\,pq}\w(\d_\aa H)_{pq}~.
\eeq
where $(\d H)_{pq} = (\d H)_{mpq} \dd x^m$ with obvious generalisations to other forms. We can solve this as a formal series in $\ap$. To get some intuition, write out the first few terms:
\beq\notag
\begin{split}
 \d_\aa H &\= \frac{\ap}{2}\tr(\aa\w F) + \frac{\ap}{4}R^{\H\,pq}\w \big[  \frac{\ap}{2}\tr(\aa\w F) + \frac{\ap}{4}R^{\H\,rs}\w(\d_\aa H)_{rs}\big]_{pq}\\[6pt]
 &\= \frac{\ap}{2}\tr(\aa\w F) + \frac{\ap^2}{8}R^{\H\,pq}\w  \tr(\aa\w F)_{pq}  + \frac{\ap^3}{16}R^{\H\,pq}\w[ R^{\H\,rs} \w \tr(\aa\w F)_{rs}\big]_{pq} +\cdots~.
\end{split}
\eeq
This resembles the structure of a geometric series. To sum the series in $\ap$, note that if $\eta$ is any three-form,  we have\footnote{If $A$ is a 3-form and $B$ a 5-form, then $A_p \w B = 0$ and so $0=(A_p \w B)_q =  A_{pq} \w B + A_p \w B_q$. Similarly, $0 =  (A\w B_q)_p = A_p \w B_q + A\w B_{pq}$. Hence,  $A_{pq} \w B = A\w B_{pq}$. Then identify $A= \d_\aa H$ and $B=R^{pq} \w \eta$.
}
\beq
\label{eq:W-gauge-contraction-pairing}
 R^{\H\,pq}\w(\d_\aa H)_{pq} \w\eta \= (\d_\aa H) \w [R^{\H\,pq}\w\eta]_{pq} \= \d_\aa H\w\mathcal K(\eta)~,
\eeq
where 
\beq\label{eq:cKdef}
\begin{aligned}
\cK(\eta) &\= (R^{\H\,pq}\w\eta)_{pq}\= \frac{1}{3!}\Big[ R^{\H\,pq}{}_{pq}\eta_{mnr} - 2R^{\H\,pq}{}_{pm}\eta_{qnr} - 2R^{\H\,pq}{}_{pn}\eta_{qrm} \\[5pt]
&\qquad - 2R^{\H\,pq}{}_{pr}\eta_{qmn} + R^{\H\,pq}{}_{mn}\eta_{pqr} + R^{\H\,pq}{}_{nr}\eta_{pqm} + R^{\H\,pq}{}_{rm}\eta_{pqn}\Big]\dd x^{mnr}~.
\end{aligned}
\eeq
Wedging \eqref{eq:W-gauge-recursion} by $\eta$, 
\beq
\d_\aa H \w \eta \= \frac{\ap}{2}\tr(\aa\w F) \w \eta + \frac{\ap}{4}  \d_\aa H \w \cK(\eta)~,
\eeq
this gives 
\beq
\d_\aa H \w \left( 1 - \frac{\ap}{4} \cK \right) \eta  \=  \frac{\ap}{2}\tr(\aa\w F) \w \eta~.\\[5pt]
\eeq
Now pick $\eta$ to be the following three-form,
\beq
\label{eq:W-gauge-effective-Omega}
{\wh\Omega \=  \left(1 - \frac{\ap}{4}\mathcal K\right)^{-1}\Omega
 ~,}
\eeq
and substitute we find
\beq
  \d_\aa H \w  \O \= \frac{\ap}{2}\tr(\aa\w F) \w \wh\O~.
\eeq
Putting it all together, the variation of the superpotential is
\beq\label{eq:gauge_stationary_W}
\d_\aa W \= \int_X \d_\aa H \w \O  \= \int_X \frac{\ap}{2}\tr(\aa\w F) \w \wh \O  ~.
\eeq
For $\ap$ small, the three-form $\wh \O$ is smoothly connected to the original $\O$:
\beq
\label{eq:W-gauge-effective-Omega-series}
\wh\Omega \= \Omega + \frac{\ap}{4}\mathcal K\Omega + \frac{\ap^2}{16}\mathcal K^2\Omega + \cdots \= \sum_{j \= 0}^{\infty}\frac{\ap^j}{4^j}\mathcal K^j\Omega~.
\eeq
That being so, $\wh \Omega$ is  non-degenerate $\wh \O \w \ol {\wh\O} \ne 0$. If we assume it is decomposable in the sense of \cite{Hitchin:2000jd}---a very reasonable assumption as decomposability is open---it defines an almost complex structure on $X$. The stationary point \eqref{eq:gauge_stationary_W} involved $\aa$, which is constructed with $J$. So it is 
$[F\w\wh\Omega]^{3,2}_J = 0$.
Since $J$ and $\wh J$ are close to each other projection from $(3,2)_{\wh J}$ to $(3,2)_{J}$ is invertible and identity for $\ap=0$. Hence, the stationary point is
\beq
[F^{0,2}]_{\wh J} \= 0~.
\eeq
Note we have not yet shown that $\dd \wh \O = 0$ and so defines an integrable complex structure, so this is not yet holomorphic bundle condition. The point here is that this is an example of how we can find all order in $\ap$ equations by summing the $\ap$-expansion implicit in $H$. Expanding this equation perturbatively in $\ap$, the bundle is holomorphic
$$
[ F^{0,2}]_{J} \= \cO(\ap^3)~.
$$
This is consistent with the ten-dimensional supersymmetry transformations of BdR \cite{Bergshoeff:1989de} as we show in  \eqref{eq:W-cubic-k-gauge}.

The same iteration works for a general fluctuations, though the details are more complicated. The resulting variation is
\beq
\label{eq:W-gauge-resummed-variation_real}
\begin{aligned}
\d W & \= -\int_X\ccZ\w\dd\Omega + \int_X(H + \ii\dd\o)\w\d\Omega + \frac{\ap}{2}\int_X\tr(\aa \w F)\w\wh \Omega\\
&\quad - \ap\int_X M_{mp}\dd x^m\w\left[\nabla^-_q(R^{\H\,pq}\w\wh \Omega) + \half H_{rs}{}^pR^{\H\,rs}\w\wh \Omega\right]~.
\end{aligned}
\eeq
 In the last line, $\nabla^-$ acts on every index of $R^\H\w\wh\Omega$ before we take its complex type. 

Substituting for $M$,  we find five independent variations:
\beq
\label{eq:W-gauge-resummed-variation_hol}
\begin{aligned}
\d W & \= -\int_X\ccZ^{0,2}\w\delb\Omega + \int_X kH^{0,3}\w\Omega\\
&\quad - \half\int_X\D^\r\w(H + \ii\delb\o)_{\r\kb\lb}\dd x^\kb\w\dd x^\lb\w\Omega + \frac{\ap}{2}\int_X\tr(\aa\w F)\w\wh \Omega\\
&\quad - \ap\int_X g_{\r\nb}\D^\r\w\left[\nabla^-_q(R^{\H\,\nb q}\w\wh \Omega) + \half H_{rs}{}^\nb R^{\H\,rs}\w\wh \Omega\right]\\
&\quad - \frac{\ap}{2}\int_X\ccZ_{\tb\nb}\dd x^\tb\w\left[\nabla^-_q(R^{\H\,\nb q}\w\wh \Omega) + \half H_{rs}{}^\nb R^{\H\,rs}\w\wh \Omega\right]\\
&\quad + \frac{\ap}{2}\int_X\ccZ_{\m\tb}\dd x^\tb\w\left[\nabla^-_q(R^{\H\,\m q}\w\wh \Omega) + \half H_{rs}{}^\m R^{\H\,rs}\w\wh \Omega\right]~.
\end{aligned}
\eeq

\section{Comparison against ten-dimensional supersymmetry transformations of BdR}
\label{sec:BdR-checks}

Our goal now is to compare these stationary points with the supersymmetry equations at $\ap^2$ and $\ap^3$  derived in BdR \cite{Bergshoeff:1989de}, which at $\ap^2$ were applied to $SU(3)$ manifolds in \citeMP and  at $\ap^3$ in upcoming work \citeapthree.

\subsection{First-order in $\ap$}

Recall, that at zeroth order in $\ap$, the stationary equations gave $H\=\dd^c\o$ and so this identifies $\nabla^-$ with Bismut to this order. It therefore preserves complex type. The pair exchange identity \eqref{eq:pair} and the Bianchi identity give
\beq
\label{eq:W-derived-curvature-order}
R^\H_{\mb\nb pq}\=\cO(\ap) \quad \Longrightarrow \quad R^{\H\,0,2} \= \cO(\ap)~.
\eeq

Consequently, the terms with $R^\H$  in $\d W$ start at $\ap^2$. The stationary equations at first order in $\ap$ are 
\beq
\label{eq:W-summary-first-order}
H\=\dd^c\o+\cO(\ap^2)~, \qquad \dd\Omega \=\cO(\ap^2)~, \qquad \ap F^{0,2}\=\cO(\ap^2)~.
\eeq
We used $R^{\H\,0,2}\=\cO(\ap)$. The stationary equations of $W$ give only part of the supersymmetry conditions derived in Hull \cite{Hull:1986kz} and Strominger \cite{Strominger:1986uh}. There are two others: the conformally balanced equation and the trace part of the Hermitian Yang--Mills equation. At $\cO(\ap)$, these are
\beq
\label{eq:lower-order-Dterms}
\dd(e^{-2\Phi}\o^2) \= \cO(\ap^2)~, \qquad g^{\m\nb}F_{\m\nb} \= \cO(\ap^2)~.
\eeq
The first follows from the gravitino and dilatino equations, while the second follows from the gaugino equation; see eqs.~(2.10), (2.34) and Section~3.5 of \citeMP. In four-dimensional language, these are D-term conditions, imposed alongside the F-term conditions from $W$. We call them the lower-order D-terms because here we use their zeroth- and first-order parts. Furthermore, as shown in \cite{McOrist:2021dnd} the linearisation of  \eqref{eq:W-summary-first-order} are the kernel of a $\Dbar$ operator while the linearisation of \eqref{eq:lower-order-Dterms} lie in the kernel of its adjoint, further justification of the terminology F-terms and D-terms.

\subsection{Second-order in $\ap$}
We now expand the stationary equations \eqref{eq:W-gauge-resummed-variation_hol} to $\ap^2$. To do so, first  introduce three tensors, the first is familiar from  BdR \cite{Bergshoeff:1989de}: 
\beq
\label{eq:PCq-definitions}
\begin{aligned}
P_{mnp}&\=\frac14e^{2\Phi}(\nabla^-)^q(e^{-2\Phi}\dd H)_{qmnp}~,\\[4pt]
C_m&\=2\del_m\Phi+ \frac{\overline\Omega^{\m\n\r}\nabla^-_m\Omega_{\m\n\r}} {3!\,\|\Omega\|_g^2}~, \qquad q_H\=g^{\m\nb}g^{\r\bar\s}(\dd H)_{\m\nb\r\bar\s}~.
\end{aligned}
\eeq
The $\D$ variation gives the first equation below, which we call the flux-torsion relation. If we take the trace of this equation with respect to the hermitian metric, combined with the $\ccZ^{0,2}$ variation we find the equation for the volume form $\O$. The remaining equations follow from the $k$ equation (rescaling of $\O$) and the $\aa$ variation. Details are  in the Appendix~\ref{app:second-order}:
\beq
\label{eq:W-summary-reduction}
\begin{aligned}
\D:\quad
&(H-\dd^c\o-2\ap P)_{\r\kb\lb}\=\frac{\ap}{2}C^m(\dd H)_{m\r\kb\lb}
-\frac{\ap}{4}H_\r{}^{mn}(\dd H)_{mn\kb\lb}
+\cO(\ap^3)~,\\[5pt]
k:\quad
&H^{0,3}\=\cO(\ap^3)~,\\[5pt]
\ccZ^{0,2}:\quad
&\dd(e^{-\ap q_H/8}\Omega)\=\cO(\ap^3)~,\\[5pt]
\aa:\quad
&\ap F^{0,2}\=\cO(\ap^3)~.
\end{aligned}
\eeq
We want to compare these equations to that of BdR \cite{Bergshoeff:1989de}. The definition $\O$ gives $\|\Omega\|_g\=e^{-2\Phi}$. The first order $\ap$ F-term equations then imply
\beq
\label{eq:balance-C}
\delb(e^{-2\Phi}\o^2)\=-e^{-2\Phi}C_\lb\dd x^\lb\w\o^2+\cO(\ap^2)~.
\eeq
Thus $C$ measures the lower order conformally balanced equation \eqref{eq:lower-order-Dterms}.  The first equation in \eqref{eq:W-summary-reduction} becomes
\beq
\label{eq:W-summary-reduction_final}
{\begin{aligned}
&(H-\dd^c\o)_{\r\kb\lb} -2\ap\Big[P_{\r\kb\lb}-\frac18H_\r{}^{mn}(\dd H)_{mn\kb\lb}\Big]\\
&\qquad\=\frac{\ii\ap}{4}e^{2\Phi} \Big[\star\dd(e^{-2\Phi}\o^2)\Big]^\m (\dd H)_{\m\r\kb\lb}+\cO(\ap^3)~.
\end{aligned}}
\eeq
 In our conventions, the order-$\ap^2$ gravitino variation, (A.6) and (A.8) of \cite{Bergshoeff:1989de},  reads\footnote{The distinction between a derivative acting on Lorentz indices and one acting on all tensor indices is responsible for the $H\dd H$ term.}
\beq
\label{eq:BdR-gravitino}
\d\psi_m\=\nabla^-_m\eta+ \frac{\ap}{4}\Big[P_{mab}-\frac18H_m{}^{pq}(\dd H)_{pqab}\Big] \gamma^{ab}\eta+\cO(\ap^3)~.
\eeq
Rewriting the vanishing of this in terms of a tensor equation following \cite{Hull:1986kz,Strominger:1986uh}, see \citeMP for details at $\ap^2$, gives the left hand side of  \eqref{eq:W-summary-reduction_final}.\footnote{In \citeMP, $H = \cO(\ap)$, so $\ap H\dd H = \cO(\ap^3)$ and is dropped.}

The $k$ equation of \eqref{eq:W-summary-reduction} agrees with the $(0,3)$ part of the dilatino equation in (A.6) and (A.8) of \citeBdR. In our notation, this gives
\beq\label{eq:lowerorderH03}
H^{0,3} - 3\ap P^{0,3} \= \cO(\ap^3)~.
\eeq
This is the complex conjugate of eq.~(2.33) of \citeMP. The type argument in Section~3.2 of \citeMP gives $P^{0,3} = \cO(\ap^2)$, so this reduces to $H^{0,3} = \cO(\ap^3)$, as in \eqref{eq:W-summary-reduction}.

The volume-form equation of \eqref{eq:W-summary-reduction} uses both the gravitino and dilatino equations. Section~3.4 of \citeMP derives the required rescaling of the spinor bilinear; see eqs.~(3.22)--(3.26). After using the lower-order D-terms, their holomorphic volume form agrees with our $e^{-\ap q_H/8}\Omega$ through $\ap^2$. We spell out this identification in the next subsection.

The bundle equation is the $(0,2)$ part of the gaugino equation (A.7) of \citeBdR, whose explicit second-order correction vanishes in (A.9). The same equation, including the overall factor of $\ap$, appears in the third line of eq.~(2.10) of \citeMP. The other component of the gaugino equation gives the Hermitian Yang--Mills condition, which we impose separately as a D-term.

Finally, the $\ccZ^{1,1}$ coefficient vanishes through $\ap^2$ after using the lower-order stationary equations. It therefore gives no additional independent condition at this order.

{\bf Remark}: The difference in the flux--torsion equation \eqref{eq:W-summary-reduction_final} is proportional to the conformally balanced D-term. Adding a multiple of a D-term to an F-term equation does not change  the supersymmetry locus at this order. Appendix~\ref{app:four-dimensional} gives a simple four-dimensional example.

To summarise, the stationary points of $W$, expanded to $\ap^2$ in \eqref{eq:W-summary-reduction}, agree with the supersymmetry transformations of BdR to $\ap^2$. 

\subsection{Holomorphic volume form and field redefinitions}

We now check that our stationary equations give the field redefinitions found in \citeMP. This also identifies  $\wh\Omega$ with their holomorphic volume form through $\ap^2$.

On compact $X$ with a smooth $\ap\to0$ limit, the lower-order F-terms and D-terms give $H,\dd\Phi,P = \cO(\ap)$ and $C = \cO(\ap^2)$. The $\ap H\dd H$ term then starts at $\ap^3$, and the flux--torsion equation reduces to
\beq
\label{eq:W-summary-torsion}
H^{1,2} + \ii\delb\o - 2\ap P^{1,2} \= \cO(\ap^3)~.
\eeq
The $P$ correction can be absorbed into $\o$, as in eqs.~(1.5)--(1.6) of \citeMP:
\beq
\label{eq:W-summary-hatted-torsion}
\begin{aligned}
H &\= \dd^c\wh\o + \cO(\ap^3)~,\qquad \wh\o \= \o + \frac{\ii\ap^2}{4}\Big[\tr(F_{\m p}F_\nb{}^p) - \tr(R^\H_{\m p}R^\H_\nb{}^p)\Big]\dd x^{\m\nb} ~, \\[3pt]
h &\= \tr|F|^2 - \tr|R^\H|^2~.
\end{aligned}
\eeq
We have used both the conformally balanced and Hermitian Yang--Mills equations in reaching this result. In other words, this is the F-term locus supplemented with D-term equations. 

Next we compare the volume forms. Denote $|F|^2 = \half F_{mn}F^{mn}$ and likewise for $R^\H$. Contracting the Bianchi identity with $g^{\m\nb}$ twice gives
\beq
\label{eq:qH-Dterms}
q_H \= \frac{\ap}{2}\Big[h + \tr\big((g^{\m\nb}F_{\m\nb})^2\big) - \tr\big((g^{\m\nb}R^\H_{\m\nb})^2\big)\Big] + \cO(\ap^3)~.
\eeq
The   hermitian trace of $F$ is $\cO(\ap^2)$ by the lower-order HYM equation. For the Hull curvature its trace is,
\beq
\label{eq:Hull-trace-C}
g^{\m\nb}R^\H_{\m\nb ab} \= (\dd C)_{ab} + \half g^{\m\nb}(\dd H)_{\m\nb ab} + \cO(\ap^3)~.
\eeq
Since $C = \cO(\ap^2)$ and $\dd H = \cO(\ap)$, the trace of $R^\H$ is $\cO(\ap)$. Hence $q_H = \ap h/2 + \cO(\ap^2)$.

The pair exchange identity \eqref{eq:pair} also gives $\cK(\Omega) = -\half q_H\Omega + \cO(\ap^2)$. Substituting this into the geometric series \eqref{eq:W-gauge-effective-Omega-series}, and using the volume-form equation in \eqref{eq:W-summary-reduction}, gives
\beq
\label{eq:W-volume-MP}
{\wh\Omega \= e^{-\ap^2h/16}\Omega + \cO(\ap^3)~, \qquad \dd\wh\Omega \= \cO(\ap^3)~.}
\eeq
With our definition $\Omega = e^{-2\Phi}\Psi$, this reads $\wh\Omega = e^{-2\Phi - \ap^2h/16}\Psi + \cO(\ap^3)$. This is the holomorphic volume form in eq.~(3.26) of \citeMP: the geometric series gives the same rescaling of the spinor bilinear.

Through $\ap^2$, the stationary equations of $W$ agree with the corresponding BdR supersymmetry equations on an $SU(3)$ manifold, after imposing the lower-order D-terms. They also reproduce the field redefinitions found in \citeMP. This gives a non-trivial check of $W$: its F-term equations, together with the D-terms, describe the same supersymmetry locus as the ten-dimensional theory at this order.

\section{The $\ap^3$ stationary equations}
\label{sec:cubic}

We now expand the stationary points to $\ap^3$. BdR give the spinor equations to $\ap^3$ which completely fix the Chern--Simons gravity sector of the theory. In upcoming work \citeapthree, the analysis of \citeMP is repeated on these $\ap^3$ fermionic corrections. What we find here is that some of those equations precisely match the stationary points of $W$ computed here.  

\subsection{The flux--torsion equation}
\label{sec:cubic-flux}

Consider the $\D$ variation in \eqref{eq:W-gauge-resummed-variation_hol}. Recall, its stationary equation is:
\beq
\label{eq:W-Delta-cubic-start}
\begin{aligned}
&\half(H - \dd^c\o)_{\r\kb\lb}\,\dd x^{\kb\lb} \w\Omega\\
&\qquad \= -\ap g_{\r\nb}\Big[\nabla^-_q\big(R^{\H\,\nb q}\w\wh\Omega\big) + \half H_{rs}{}^\nb R^{\H\,rs}\w\wh\Omega\Big]^{3,2}_J~.
\end{aligned}
\eeq
The derivative acts on all tensor indices before the type projection. All complex types in this section refer to $J$.

For the comparison with the $\ap^3$ supersymmetry transformations in BdR, the calculation is significantly more complicated. As BdR and \citeapthree are onshell calculations, we can, from the outset, use the first-order  $\ap$ D-terms in \eqref{eq:lower-order-Dterms} to simplify intermediate expressions. These are extra conditions on the background, imposed after taking the variation. At $\ap$, D-terms with their corresponding F-terms, compactness and a smooth $\ap\to0$ limit, imply \citeMP:
\beq
H,\ \dd\Phi,\ P \= \cO(\ap)~, \qquad C \= \cO(\ap^2)~.
\eeq
The estimate for $C$ uses conformally balanced. In particular, the term $\ap C^m(\dd H)_{m\r\kb\lb}$ in \eqref{eq:W-summary-reduction} is now $\cO(\ap^4)$.

From \eqref{eq:W-volume-MP}, we  have
\beq
\wh\Omega \= \left(1 - \frac{\ap^2h}{16}\right)\Omega + \cO(\ap^3)~.
\eeq
Since  $R^\H\w \O = \cO(\ap)$, the corresponding curvature term in \eqref{eq:W-Delta-cubic-start} starts at $\ap^2$ and  the above rescaling contributes at $\ap^4$. This also holds when the derivative acts on the scalar $h$. The remaining $\cO(\ap^3)$ correction is multiplied by the explicit $\ap$, so it first contributes at $\ap^4$ as well. Hence, we can replace $\wh\Omega$ by $\Omega$ in \eqref{eq:W-Delta-cubic-start}.  We must, however, keep the modification to the minus connection supplied by the lower-order flux--torsion equation \eqref{eq:lowerorderH03}--\eqref{eq:W-summary-torsion}:
\beq\label{eq:minusconnection}
\Gamma^- \= \Gamma^{\mathrm B} - \ap P + \cO(\ap^3)~,
\eeq
where $\Gamma^{\mathrm B}$ is the Bismut connection of $(g,J)$. Substituting \eqref{eq:minusconnection} into \eqref{eq:W-Delta-cubic-start} and using the pair exchange identity \eqref{eq:pair} gives (see \sref{s:fluxtorsionap3} for more details):
\beq
\label{eq:W-Delta-MP-cubic}
\begin{aligned}
&(H - \dd^c\o - 2\ap P)_{\r\kb\lb} + \frac{\ap}{4}H_\r{}^{mn}(\dd H)_{mn\kb\lb}\\
&\qquad \= \ap^2(\nabla^-)^q\left(\nabla^-_\r P_{q\kb\lb} - \nabla^-_qP_{\r\kb\lb}\right) + \cO(\ap^4)~.
\end{aligned}
\eeq
This is the extension of the flux torsion F-term \eqref{eq:fluxtorsionzero}  through $\ap^3$. When compared with the $\ap^2$ result, \eqref{eq:W-summary-reduction_final}, the left hand side is unchanged, while the right hand side is the new $\ap^3$ correction.

We can check this contribution directly against BdR's eq.~(A.10) \citeBdR. With their $\alpha = \beta = -\ap/4$ and $T_{mnpq} = -(\dd H)_{mnpq}/6$, we have $\nabla^qT_{qmab} = -2P_{mab}/3 + \cO(\ap^2)$. The first term in their $\ap^3$ gravitino correction becomes
\beq
\label{eq:BdR-cubic-flux-term}
\frac{\ap^2}{8}\nabla^q\left(\nabla_mP_{qab} - \nabla_qP_{mab}\right)\gamma^{ab}\eta + \cO(\ap^4)~.
\eeq
 The remaining curvature terms in (A.10) of \citeBdR act on $\eta$ as scalar multiples of $\eta$ at $\ap^3$.  This follows from  their Clifford products having equal numbers of holomorphic and antiholomorphic indices. They therefore do not contribute to the flux--torsion equation. 
 
 Returning to \eqref{eq:BdR-cubic-flux-term}, this is to be added to \eqref{eq:BdR-gravitino} and we see it plays the role of deforming the torsion of $\nabla^-$. Since torsion enters $\nabla^-_m\eta$ as $-H_{mab}\gamma^{ab}\eta/8$, \eqref{eq:BdR-cubic-flux-term} gives the right hand side of \eqref{eq:W-Delta-MP-cubic}, including its coefficient and sign. 
 
 Hence, the supersymmetry transformations of \citeBdR match the stationary point of $W$ up to D-terms.  This is another non-trivial check of both the superpotential as a functional reproducing supersymmetry equations and the equations of BdR. 

\subsection{Field redefinitions revisited}
In \citeMP, we found a convenient field redefinition of $\o$ and $\Phi$ that absorbed almost all of the explicit $\ap^2$ corrections to the F-terms, and in fact, D-terms. Here we briefly comment on a extension of this field redefinition to $\ap^3$  that can simplify the flux torsion F-term \eqref{eq:W-Delta-MP-cubic}.  

Define, 
\beq
\cE^\H \= \Big[\tr(F_{\m p}F_\nb{}^p) - \tr(R^\H_{\m p}R^\H_\nb{}^p)\Big]\dd x^{\m\nb}~,
\eeq
and define its contraction with the curvature by
\beq
R\cdot\cE^\H \= R_{\r\kb}{}^{\m\nb}\cE^\H_{\m\nb}\,\dd x^{\r\kb} ~.
\eeq

Using the trace identity \eqref{eq:Hull-trace-C} with $C = \cO(\ap^2)$ and tracing the Bianchi identity \eqref{eq:pair} once with the hermitian metric, gives
\beq
\tr\left((g^{\s\tb}R^\H_{\s\tb})R^{\H\,1,1}\right) \= -\frac{\ap}{2}R\cdot\cE^\H + \cO(\ap^2)~.
\eeq
Substitute this into the definition of  $P$ in \eqref{eq:PCq-definitions}, simplify with the lower-order D-terms \eqref{eq:lower-order-Dterms}: 
\beq
\label{eq:W-Delta-P-cubic}
\begin{aligned}
2\ap P_{\r\kb\lb} &\= \left\{\delb\left[\frac{\ap^2}{4}\cE^\H + \frac{\ap^3}{8}R\cdot\cE^\H\right]\right\}_{\r\kb\lb} + \frac{\ap}{2}H_\r{}^{mn}(\dd H)_{mn\kb\lb} + \cO(\ap^4)~.
\end{aligned}
\eeq
The $\cO(\ap^2)$ remainders in the D-term equations first affect this expression at $\ap^4$. For \eqref{eq:BdR-cubic-flux-term}, we only need the lower order expression:
\beq
P^{1,2} \= \frac{\ap}{8}\delb\cE^\H + \cO(\ap^2)~, \qquad \nabla^qP_{qab} \= \cO(\ap^2)~.
\eeq
Splitting the sum in \eqref{eq:BdR-cubic-flux-term} over $q$ into holomorphic and antiholomorphic parts gives a Laplacian and a curvature commutator. The \K identities and curvature Bianchi identity then give
\beq
\label{eq:W-Delta-divergence-cubic}
\begin{aligned}
&\ap^2\nabla^q\left(\nabla_\r P_{q\kb\lb} - \nabla_qP_{\r\kb\lb}\right)\\
&\qquad \= -\frac{\ap^3}{8}\left\{\delb\left[\nabla^q\nabla_q\cE^\H + 2R\cdot\cE^\H\right]\right\}_{\r\kb\lb} - \frac{\ap^2}{2}\big[\tr(R\w P)\big]_{\r\kb\lb} + \cO(\ap^4)~.
\end{aligned}
\eeq
Inside the trace, $R$ and $P$ are Lorentz-valued forms:
\beq
R^a{}_b \= \half R_{mn}{}^a{}_b\dd x^m\w\dd x^n~, \qquad P^a{}_b \= P_m{}^a{}_b\dd x^m~.
\eeq
The zeroth order \K curvature types and the antisymmetry of $P_{mab}$ give
\beq
\big[\tr(R\w P)\big]_{\r\kb\lb} \= -4R_{\r[\kb}{}^{\m\nb}P_{\m|\nb|\lb]} + \cO(\ap^2)~,
\eeq
where antisymmetrisation includes a factor of $1/2$.

We can absorb both $\delb$ derivatives  in \eqref{eq:W-Delta-divergence-cubic} into a shift of $\o$:
\beq
\label{eq:W-Delta-omega-cubic}
{\wh\o \= \o + \frac{\ii\ap^2}{4}\cE^\H - \frac{\ii\ap^3}{8}\left(\nabla^q\nabla_q\cE^\H + R\cdot\cE^\H\right)~.}
\eeq
The first correction is the shift found in \citeMP; the second is its $\ap^3$ extension. This redefinition keeps $J$ fixed. With $\dd^c = \ii(\del - \delb)$, it absorbs the derivative terms above and their complex conjugates.

Substituting into \eqref{eq:W-Delta-MP-cubic}, the $H\dd H$ terms combine to give
\beq
\label{eq:W-Delta-hatted-cubic}
{\begin{aligned}
(H - \dd^c\wh\o)_{\r\kb\lb} &\= \frac{\ap}{4}H_\r{}^{mn}(\dd H)_{mn\kb\lb} - \frac{\ap^2}{2}\big[\tr(R\w P)\big]_{\r\kb\lb} + \cO(\ap^4)~.
\end{aligned}}
\eeq
Both terms on the right start at $\ap^3$. To understand these two terms, first note that  $\Th^\H - \Th^\Ch = \cS+\overline{\cS}+\ap P+\cO(\ap^3)$, where
\beq
\cS^\n{}_\sb \= (H-2\ap P)_\m{}^\n{}_\sb\,\dd x^\m~.
\eeq
Up to an exact form, their Chern--Simons forms differ by 
\beq
\frac{\ap}{4} \left( CS(\Th^\H)- CS(\Th^\Ch)\right) \= \frac{\ap}{2}  \cS^\n{}_\sb\w\dd_\nabla\overline{\cS}^{\sb}{}_\n
+\frac{\ap^2}{2} \tr(P\w R)+\cO(\ap^4)~,
\eeq
where $\dd_\nabla$ uses the zeroth order connection on a Calabi--Yau. Using $H=\dd^c\o+\cO(\ap^2)$ gives
\beq
\begin{aligned}
&\frac{\ap}{4}
\left[\operatorname{CS}(\Th^\Ch)
-\operatorname{CS}(\Th^\H) + \delb\,\tr(\cS\w\overline{\cS})\right]_{\r\kb\lb}\\
&\qquad\=
\frac{\ap}{4}H_\r{}^{mn}(\dd H)_{mn\kb\lb}
-\frac{\ap^2}{2}\big[\tr(R\w P)\big]_{\r\kb\lb}+\cO(\ap^4)~.
\end{aligned}
\eeq
Hence, the $\ap^3$ F-term \eqref{eq:W-Delta-MP-cubic} can be simplified by a field redefinition to
\beq
\label{eq:W-Delta-hatted-cubic_b}
\begin{aligned}
&\left[
H+i\delb \left( \wh\o
+\frac{\ii\ap}{4}\tr(\cS\w\overline{\cS})\right)
\right]_{\r\kb\lb}\\
&\qquad\=
\frac{\ap}{4}
\left[
\operatorname{CS}(\Th^\Ch)-\operatorname{CS}(\Th^\H)
\right]_{\r\kb\lb}
+\cO(\ap^4)~.
\end{aligned}
\eeq
where the Chern connection is computed with respect to $\wh \o$.

\subsection{The $\aa$ and $k$ stationary equations}

We now simplify $k$ and $\aa$   stationary equations in \eqref{eq:W-gauge-resummed-variation_hol} to $\ap^3$ for comparison with BdR using the same lower-order F-terms and D-terms as above. The $k$ variation, nothing changes.  For the gauge variation $\aa$, \eqref{eq:W-volume-MP} gives $\wh\Omega = (1 - \ap^2h/16)\Omega + \cO(\ap^3)$. Thus
\beq
\label{eq:W-cubic-k-gauge}
{\begin{aligned}
k:\quad H^{0,3} &\= \cO(\ap^4)~,\\[4pt]
\aa:\quad \ap F^{0,2} &\= \cO(\ap^4)~.
\end{aligned}}
\eeq
Both types refer to $J$.

These agree with the corresponding BdR equations at $\ap^3$. Indeed, using the $\ap^2$ results in \citeMP, $P^{1,2} = \ap\delb\cE^\H/8 + \cO(\ap^2)$ gives $(\dd H)^{1,3} = \cO(\ap^3)$ and hence $P^{0,3} = \cO(\ap^3)$. The  $\ap^3$ terms in eq.~(A.10) of \citeBdR have no  $(0,3)$ contributions.  The correction to the  gaugino variation in eq.~(A.11) starts at $\ap^3$, so its contribution to our  gauge equation starts at $\ap^4$. As stated earlier on, we will report on these $\ap^3$ results in more detail in upcoming work \citeapthree.

\subsection{The $\ccZ^{0,2}$ stationary equation}
Combining the $\ccZ^{0,2}$ equation with the hermitian trace of the $\D$ equation \eqref{eq:W-Delta-cubic-start} gives
\beq
\label{eq:W-cubic-volume-trace}
\delb\Omega \= \half g^{\r\nb}(H - \dd^c\o)_{\r\nb\lb}\,\dd x^\lb\w\Omega + \cO(\ap^4)~,
\eeq
which will be the extension of the third line of \eqref{eq:W-summary-reduction} to $\ap^3$.
To see this, first simplify this equation by taking the hermitian trace of \eqref{eq:W-Delta-MP-cubic} and substitute the result.

We need to expand $\wh \O$ to $\ap^3$.  To do this we use start with the geometric series \eqref{eq:W-gauge-effective-Omega-series} for $\wh \Omega$ and use the following identities:
\beq
\label{eq:W-cubic-volume-contractions}
q_H \= \frac{\ap}{2}h + \cO(\ap^3)~, \qquad g^{\r\nb}P_{\r\nb\lb} \= \frac{\ap}{16}\del_\lb h + \cO(\ap^2)~.
\eeq
Using these substitutions and the pair exchange identity \eqref{eq:pair} as described in \sref{s:vol_form}, the geometric series \eqref{eq:W-gauge-effective-Omega-series} becomes:
\beq
\label{eq:W-cubic-volume-expansion}
\begin{aligned}
\wh\Omega &\= \left(1 - \frac{\ap^2h}{16} + \frac{\ap^3}{32}\nabla^q\nabla_q h\right)\Omega - \frac{\ap^3}{32}\dd\left[(\nabla^\r h)\Omega_\r\right] + \cO(\ap^4)~.
\end{aligned}
\eeq
Thus, at $\ap^3$, $\wh\Omega$  has a $(2,1)$ component but it is $\dd$--exact. 

The $\ccZ^{0,2}$ variation sets $(\dd\wh\Omega)^{3,1} = \cO(\ap^4)$, and combined with  \eqref{eq:W-cubic-volume-expansion} we obtain
\beq
\label{eq:W-cubic-volume}
\dd\wh\O \= \dd\left[e^{-\ap^2h/16 + \ap^3\nabla^q\nabla_q h/32}\Omega\right] \= \cO(\ap^4)~.
\eeq

To compare with BdR, the $\ap^3$ corrections in (A.8) and (A.10) cancel in the combination $\d\lambda + \sqrt{2}\gamma^m\d\psi_m/4$. Rewriting its one-gamma component using $\Omega = e^{-2\Phi}\Psi$ gives \eqref{eq:W-cubic-volume-trace}. Together with the flux--torsion equation \eqref{eq:W-Delta-MP-cubic} this gives \eqref{eq:W-cubic-volume}.  The curvature terms that act as scalars on the spinor $\d\lambda + \sqrt{2}\gamma^m\d\psi_m/4$ cancel.

\subsection{The Hermitian equation}

The $\ccZ^{1,1}$ coefficient gives no further condition through $\ap^3$. Its $P^{0,3}$ term is multiplied by $\ap$, so it starts at $\ap^4$. The $H\dd H$ contraction has no cubic contribution with three barred free indices. The remaining double derivative contains $\ap^2\nabla^q\nabla_\rb P_{q\kb\lb}$. At leading order the relevant barred derivatives commute, and $\nabla^qP_{qab} = \cO(\ap^2)$ makes this term $\cO(\ap^4)$ too.

\subsection{Summary}
Together with \eqref{eq:W-Delta-hatted-cubic}, these give the stationary equations through $\ap^3$ after using the lower-order D-terms:
\beq
\label{eq:W-cubic-stationary-summary}
\begin{aligned}
\D:\quad
&(H-\dd^c\wh\o)_{\r\kb\lb} \= \frac{\ap}{4}\left[\operatorname{CS}(\Th^\Ch)-\operatorname{CS}(\Th^\H)\right]_{\r\kb\lb} +\cO(\ap^4)~,\\[6pt]
k:\quad
&H^{0,3}_J \= \cO(\ap^4)~,\\[6pt]
\ccZ^{0,2}:\quad
&\dd\wh\Omega
\= \dd\left[e^{-\ap^2h/16+\ap^3\nabla^q\nabla_qh/32}\Omega\right]
\= \cO(\ap^4)~,\\[6pt]
\aa:\quad
&\ap F^{0,2}_J \= \cO(\ap^4)~.
\end{aligned}
\eeq
Here $\wh\o$ is defined in \eqref{eq:W-Delta-omega-cubic}, the first line is defined up to a $\delb$-exact term. An alternative expression, which is exact, is given in \eqref{eq:W-Delta-hatted-cubic}. The $\ccZ^{1,1}$ equation gives no additional condition at this order. These equations agree with the corresponding BdR equations. The cubic corrections to the D-terms we have not yet determined.

{\bf Remark}: one might wonder if the corrections on the right hand side to the $\D$ equation could be absorbed by a field redefinition. At fixed $J$ and $H$, absorbing the cubic remainder in \eqref{eq:W-Delta-hatted-cubic} requires it to be $-\ii\delb\sigma$ for a real, globally defined, local $(1,1)$-form $\sigma=\cO(\ap^3)$, which does not appear to be the case here.

\section{Holomorphy through $\ap^3$}
\label{s:holomorphy}
We now check that the antiholomorphic variation of $W$ vanishes through $\ap^3$. We use the lower-order F-terms on the background, but the fluctuations remain arbitrary: they need not satisfy the linearised stationary equations. We do not use any D-terms.

Let $\bar\a$ label an antiholomorphic direction in field space. In holomorphic parameter gauge,
\beq
\d_{\bar\a}\Omega \= 0~, \qquad (\d_{\bar\a}A)^{0,1} \= 0~, \qquad \ccZ_{\bar\a} \= \ccZ_{\bar\a}^{2,0}~, \qquad M_{\mb p} \= 0~.
\eeq
All complex types refer to $J$. We start directly from \eqref{eq:superpotential-variation}. The term containing $\d_{\bar\a}\Omega$ vanishes. Since $\dd\Omega$ has type $(3,1)$, the term $\ccZ_{\bar\a}\w\dd\Omega$ also vanishes. The direct gauge term has too many holomorphic indices, since $\d_{\bar\a}A$ has type $(1,0)$. This leaves just the Hull-connection variation:
\beq
\d_{\bar\a}W \= -\frac{\ap}{2}\int_X\tr\left[(\d_{\bar\a}\Th^\H)^{0,1}\w(R^\H)^{0,2}\right]\w\Omega~.
\eeq
We therefore only need to check the order of this pairing.

The lower-order flux--torsion equation gives $H - \dd^c\o = \cO(\ap^2)$, so $\nabla^-$ differs from Bismut by $\cO(\ap^2)$. Since Bismut preserves complex type, the pair exchange identity \eqref{eq:pair} gives
\beq
R^\H_{\kb\lb pq} \= \half(\dd H)_{\kb\lb pq} + \cO(\ap^2) \= \cO(\ap)~.
\eeq
The last equality uses the Bianchi identity. Thus $(R^\H)^{0,2} = \cO(\ap)$.

Now consider the variation of $\Th^\H$ in \eqref{eq:variation}. Since $M_{\kb p} = 0$, its terms with its form index being antiholomorphic  can only come from the part of $\nabla^-$ that changes complex type. They are therefore $\cO(\ap^2)$, while the explicit $H$ term vanishes. Hence
\beq
(\d_{\bar\a}\Th^\H)_{\kb pq} \= \frac{\ap}{4}\left[\tr\left(\d_{\bar\a}A\w F - \d_{\bar\a}\Th^\H\w R^\H\right)\right]_{\kb pq} + \cO(\ap^2)~.
\eeq
At zeroth order, this equation says that $\d_{\bar\a}\Th^\H$ has form type $(1,0)$. The zeroth order curvatures $F$ and $R^\H$ have type $(1,1)$, so the three-form on the right has type $(2,1)$ through $\ap$. With its first index barred, the other two indices must therefore be holomorphic. We obtain
\beq
\begin{aligned}
(\d_{\bar\a}\Th^\H)_{\kb\m\n} &\= \cO(\ap)~,\\
(\d_{\bar\a}\Th^\H)_{\kb\m\nb} &\= \cO(\ap^2)~, \qquad (\d_{\bar\a}\Th^\H)_{\kb\mb\nb} \= \cO(\ap^2)~.
\end{aligned}
\eeq

The first component needs a little care: it starts at $\ap$, but the curvature component it pairs with starts at $\ap^2$. Indeed, raising the two tangent indices exchanges holomorphic and antiholomorphic types, and \eqref{eq:pair} gives
\beq
R^\H_{\kb\lb\mb\nb} \= R^-_{\mb\nb\kb\lb} \= \cO(\ap^2)~.
\eeq
Here the $\dd H$ term vanishes because it has four barred indices on a complex threefold. The remaining estimate again follows because Bismut preserves complex type and $\nabla^-$ differs from it by $\cO(\ap^2)$.

Thus the order-$\ap$ part of the variation $\d \Th^\H$ pairs with an order-$\ap^2$ curvature component. The other components of the variation are already $\cO(\ap^2)$ and pair with $(R^\H)^{0,2} = \cO(\ap)$. In both cases,
\beq
\label{eq:W-holo-curvature-cubic}
\tr\left[(\d_{\bar\a}\Th^\H)^{0,1}\w(R^\H)^{0,2}\right] \= \cO(\ap^3)~.
\eeq
The extra factor of $\ap$ in the variation of $W$ then gives
\beq
\label{eq:W-holomorphy-cubic}
\d_{\bar\a}W \= \cO(\ap^4)~.
\eeq

We can also read off the equation for $\wh\Omega$ from a $B$-field variation. Keeping $g$, $A$, $\Omega$ and the vielbein fixed, the recursion relation for $H$ gives
\beq
\label{eq:W-gauge-effective-B-variation}
\d_BW
\= \int_X\dd\cB\w\wh\Omega
\= -\int_X\cB\w\dd\wh\Omega~.
\eeq
The derivation, including the relation to the Lorentz frame used above, is given in Appendix~\ref{app:closure}.

Once we impose the holomorphic stationary equations through $\ap^3$, the B-field stationary point follows as $\d_\cB W = 0$ is equivalent to $\d_\ccZ W = 0$ at least to $\ap^4$. Since $\cB$ is arbitrary, this gives
\beq
\label{eq:W-holo-closure-cubic}
\dd\wh\Omega \= \cO(\ap^4)~.
\eeq

Our check of holomorphy used the lower-order F-terms on the background, while leaving the fluctuations arbitrary. We have not yet shown that $W$ is holomorphic in an off-shell neighbourhood of a stationary point, or to all orders in $\ap$. These questions remain open. As the question of holomorphy is closely tied to the offshell holomorphy of deformations of the Hull connection, a possible interesting direction may be to study the lift of heterotic to its universal bundle \citeUG, extended to $\ap^2$ in \citeUGAP, in which holomorphic of such deformations are rewritten in terms of holomorphy of a tangent bundle curvature on the total space of the universal bundle. 

\section{All order predictions}
\label{sec:all-orders}

We now use \eqref{eq:W-gauge-resummed-variation_hol} without truncating its $\ap$ expansion. This gives all order predictions of $W$ in the original fields. The corrected form $\wh\Omega$ is still defined by the geometric series \eqref{eq:W-gauge-effective-Omega-series}, whose inverse is
\beq
\label{eq:W-k-all-orders-inverse}
\Omega \= \wh\Omega-\frac{\ap}{4}\mathcal K(\wh\Omega)~.
\eeq
All fields, including the curvature inside $\mathcal K$, retain their full $\ap$ dependence.

To interpret some of the equations geometrically, we assume that $\wh\Omega$ remains decomposable, with $\wh\Omega\w\overline{\wh\Omega}\ne0$. It defines an almost complex structure $\wh J$, which agrees with $J$ at $\ap=0$. 

The rescaling equation is the simplest. Varying $\Omega$ by $k\Omega$ leaves $g$, $A$ and $B$ fixed, so  $\d_kH=0$. Since $\dd\o\w\Omega=0$ on the complex background, the $k$ equation is just $H^{0,3}_J=0$. We can also write this equation using $\wh J$. Insert \eqref{eq:W-k-all-orders-inverse} into the $k$ variation and use  \eqref{eq:W-gauge-contraction-pairing} to move $\mathcal K$ off $\wh\Omega$. This gives
\beq
\label{eq:W-k-all-orders-corrected}
\left[
H+\ii\dd\o
-\frac{\ap}{4}R^{\H\,pq}\w(H+\ii\dd\o)_{pq}
\right]^{0,3}_{\wh J}
\=0~.
\eeq
This is the same rescaling equation, expressed in the corrected complex structure.  We keep the shorter form $H^{0,3}_J=0$ in the final system below.

The gauge equation has a simpler interpretation in $\wh J$. The $\aa$ variation sets $[F\w\wh\Omega]^{3,2}_J=0$. Since $\wh\Omega$ is decomposable, $F\w\wh\Omega$ has type $(3,2)$ in $\wh J$. Projection from this space to $(3,2)_J$ is the identity at $\ap=0$, so it has a perturbative inverse. Hence
\beq
\label{eq:W-gauge-all-orders}
[F\w\wh\Omega]^{3,2}_J=0
\quad\Longleftrightarrow\quad
F^{0,2}_{\wh J}=0~.
\eeq

For the volume form, the pure $B$-field variation \eqref{eq:W-gauge-effective-B-variation} gives $\d_BW=-\int_X\cB\w\dd\wh\Omega$. This follows if we assume $W$ is holomorphic to all orders in $\ap$, a very reasonable assumption given the success we have had so far in studying $W$, but we plan to return to proving this in future work. Under this assumption, then $\dd\wh\Omega=0$ and $\wh\Omega$ defines an integrable complex structure $\wh J$. Then, the $\ccZ^{0,2}$ variation vanish identically as follows directly from \eqref{eq:W-Z02-recursion}. Closure of $\wh\O$ also makes the bundle holomorphic to all orders in $\ap$; not a surprising result but reassuring the construction is consistent. 

The remaining equations come from the $\ccZ^{1,1}$ and $\D$ coefficients in \eqref{eq:W-gauge-resummed-variation_hol}. We keep them in the original complex structure $J$, with $\Omega$ on the left of the flux--torsion equation. In each curvature term, $\nabla^-$ acts on every index before taking the $(3,2)_J$ projection. Writing the flux--torsion equation entirely in $\wh J$ would also require a compatible Hermitian form and its relation to $\o$.

With the decomposability assumption stated above, the  stationary equations together with arbitrary $B$-field stationarity are therefore
\beq
\label{eq:W-all-orders-summary}
{
\begin{aligned}
k:\quad
&H^{0,3}_J \= 0~,\\[5pt]
\aa:\quad
&F^{0,2}_{\wh J} \= 0~,\\[5pt]
B:\quad
&\dd\wh\Omega \= 0~,\\[5pt]
\ccZ^{1,1}:\quad
&\Big[
\nabla^-_q\big(R^{\H\,\m q}\w\wh\Omega\big)
+\half H_{rs}{}^\m R^{\H\,rs}\w\wh\Omega
\Big]^{3,2}_J \= 0~,\\[5pt]
\D^\r:\quad
&\half(H-\dd^c\o)_{\r\kb\lb}
\dd x^\kb\w\dd x^\lb\w\Omega\\
&\qquad\= -\ap g_{\r\nb}
\Big[
\nabla^-_q\big(R^{\H\,\nb q}\w\wh\Omega\big)
+\half H_{rs}{}^\nb R^{\H\,rs}\w\wh\Omega
\Big]^{3,2}_J~.
\end{aligned}
}
\eeq

\section{Conclusion}
\label{sec:conclusion}
The goal of this paper was simple: compute the stationary loci of $W$, taking into account the gravitational Chern--Simons term in $H$. This term contains the composite Hull connection used in \citeBdR to express the supersymmetry transformations in a convenient form, and that connection itself depends on $H$. What was useful here was that we had the $\ap^2$ and $\ap^3$ results of \citeBdR, applied to SU(3) in \citeMP to compare against.  We found that the F-term equations agree with the corresponding ten-dimensional supersymmetry equations after imposing the lower-order D-terms. Some of these corrections can also be absorbed by field redefinitions. The equations are highly non-trivial, especially at $\ap^3$, so this agreement is good evidence that $W$ is a useful functional for computing how $\ap$ corrections modify the tensor equations on the compactification manifold $X$.

We have left several questions for further work. We have not yet proved that $W$ is a holomorphic functional on the space of off-shell $SU(3)$ structures, and we plan to return to this in \citeapthree.

It would also be nice to understand the role of field redefinitions, both at $\ap^3$ and to all orders in $\ap$. For example, it is striking that the F-term \eqref{eq:W-Delta-hatted-cubic_b} can be written in terms of a pair of Chern--Simons form. Taking $\dd$ of this equation, using the Bianchi identity we find a resolved Bianchi identity 
\beq
\label{eq:ddbaromega-cubic-redefined}
\begin{split}
2\ii\delb\del\,\Bigl(
\wh\o + \frac{\ii\ap}{4}\tr\left(\cS\w\overline{\cS}\right)\Bigr) &\= \frac{\ap}{4}\left(\tr F\w F - \tr R^\Ch\w R^\Ch
\right) + \cO(\ap^4)~.
\end{split}
\eeq
We will discuss this equation in \citeapthree. For now, we just note that it is manifestly of type $(2,2)$ through $\ap^3$, even though the original Bianchi identity \eqref{eq:BianchiH} is written in terms of the Hull connection.

It would also be interesting to connect the recursive structure discussed here with the approach suggested in \cite{Coimbra:2014geometry}. Using generalised geometry, the Hull connection and a Bismut identity, that work outlines another way to generate higher-order corrections to the supersymmetry equations. This might also give a way of computing how the D-terms are renormalised in $\ap$. Furthermore, our comparison concerns the corrections generated by the Yang--Mills and Lorentz Chern--Simons terms. As pointed out already in \citeBdR there is a purely gravitational multiplet that starts at $\ap^3\zeta(3)$. This term contributes to the \K potential as noted by \citeAQS, but does it affect the superpotential?

\vskip0.5cm
\paragraph*{Acknowledgements.}
I would like to thank A. Ashmore, J Knapp, R Minasian, S. Picard, S Sethi and E. Svanes  for enlightening conversations and very helpful comments. JM is supported in part by ARC Discovery Project Grants DP240101409 and DP250101828.

\appendix

\section{Further details of the variations}
\label{app:variation}

We give the intermediate steps leading to the second-order stationary equations \eqref{eq:W-summary-reduction}. We also explain the derivative convention used in the BdR comparison and the two-form variation needed for the all-order discussion. Throughout, the fluctuations are arbitrary. We use the background equations only after taking the variation.

\subsection{The connection recursion}

We start with the anomaly-covariant variation of $H$,
\beq
\d H \= \dd\cB + \frac{\ap}{2}\tr\left(\d A\w F-\d\Th^\H\w R^\H\right)~.
\eeq
Substituting the connection variation \eqref{eq:variation}, and using $M_{mp}=\half(\d g_{mp}+\cB_{mp})$, gives
\beq
\label{eq:W-gauge-complete-H-variation}
\begin{aligned}
\d H
&\= \dd\cB+\frac{\ap}{2}\tr(\d A\w F)\\
&\quad+\ap\,\dd x^m\w R^{\H\,pq}
\left(\nabla^-_qM_{mp}-\half H_{pq}{}^rM_{mr}\right)\\
&\quad+\frac{\ap}{4}R^{\H\,pq}\w(\d H-\dd\cB)_{pq}~.
\end{aligned}
\eeq
Thus the quantity being iterated is the pair of Chern--Simons forms $\d H-\dd\cB$. Using the contraction identity \eqref{eq:W-gauge-contraction-pairing}, each iteration transfers another factor of $\ap\cK/4$ onto $\Omega$. The sum is $\wh\Omega$, as in \eqref{eq:W-gauge-effective-Omega-series}. The  $\dd\cB$ term still pairs with $\Omega$. This explains why $\wh\Omega$ appears in the gauge and curvature terms of the variation, while the terms without curvature retain $\Omega$.

Integrating the derivative of $M$ by parts gives \eqref{eq:W-gauge-resummed-variation_real}. To extract the holomorphic coefficients, we use
\beq
M_{\m p}\=0~,\qquad
M_{\tb\m}\=-\half\ccZ_{\m\tb}~,\qquad
M_{\tb\nb}\=g_{\r\nb}\D^\r{}_\tb+\half\ccZ_{\tb\nb}~.
\eeq
Substitution gives \eqref{eq:W-gauge-resummed-variation_hol}. Its five independent fluctuations are $k$, $\aa$, $\D$, $\ccZ^{0,2}$ and $\ccZ^{1,1}$.

\subsection{The $\ap^2$ stationary equations}
\label{app:second-order}
The goal here is to give details of how to derive \eqref{eq:W-summary-reduction} from the all--order variation \eqref{eq:W-gauge-resummed-variation_hol} putting the stationary equations in a form suitable for comparison with BdR \citeBdR.  

\subsubsection{The $\D$  stationary equation}
The $\D$ equation gives the $\ap^2$ correction to the flux--torsion equation $H=\dd^c\o + \cO(\ap^2)$. 
The main exercise is to simplify the last three lines of \eqref{eq:W-gauge-resummed-variation_hol} involving $R^\H$. The $\D$ variation will then give 
\beq
(H-\dd^c\o)_{\r\kb\lb}
\=2\ap P_{\r\kb\lb}
+\frac{\ap}{2}C^m(\dd H)_{m\r\kb\lb}
-\frac{\ap}{4}H_\r{}^{mn}(\dd H)_{mn\kb\lb}
+\cO(\ap^3)~.
\eeq

Let's start with the $\D$ varation written in \eqref{eq:W-summary-reduction}. We will need to use  the first-order equations \eqref{eq:W-summary-first-order} to derive some preliminary results. Since $H-\dd^c\o=\cO(\ap^2)$, 
the minus connection agrees with Bismut through first order, so $R^-_{pq\lb\sb}=\cO(\ap^2)$. Using this with pair exchange \eqref{eq:pair}  gives
\beq
\label{eq:W-appendix-curvature-second-order}
R^\H_{\kb\lb pq}
\=R^-_{pq\kb\lb}+\half(\dd H)_{\kb\lb pq}
\=\half(\dd H)_{\kb\lb pq}+\cO(\ap^2)~.
\eeq
Here $\dd H=\cO(\ap)$, but we have not assumed that $H=\cO(\ap)$. As $H-\dd^c\o = \cO(\ap^2)$, with $\dd \dd^c\o$ being type $(2,2)$, it follows that $(\dd H)^{1,3}=\cO(\ap^2)$. Consequently, from \eqref{eq:cKdef} we find 
\beq
\label{eq:W-appendix-Omega-second-order}
\cK(\Omega)\=-\half q_H\Omega+\cO(\ap^2)~,\qquad
\wh\Omega\=\left(1-\frac{\ap}{8}q_H\right)\Omega+\cO(\ap^3)~,
\eeq
where the scalar comes from taking the hermitian trace of $\dd H$ twice in \eqref{eq:W-appendix-curvature-second-order}. The other complex types first contribute to $\cK\Omega$ at $\ap^2$. Since $\cK=\cO(1)$ and $\cK\Omega=\cO(\ap)$, further terms in the geometric series cannot contribute through $\ap^2$. In particular, replacing $\wh\Omega$ by $\Omega$ in the gauge and curvature terms does not change their coefficients through $\ap^2$. 

To evaluate the derivative of $\Omega$, we use the definition of $C$ in \eqref{eq:PCq-definitions}:
\beq
(\nabla^-_q\Omega)^{3,0}
\=\left(C_q-2\del_q\Phi\right)\Omega~.
\eeq
The components of $\nabla^-\Omega$ that change type are $\cO(\ap^2)$ and contribute only beyond the order needed here. 

The $\D$ equation then becomes
\beq
\label{eq:W-appendix-flux-curvature}
\begin{aligned}
(H-\dd^c\o)_{\r\kb\lb}
&\=-\ap\left((\nabla^-)^q+C^q-2\del^q\Phi\right)
R^\H_{\kb\lb\r q}\\
&\quad-\frac{\ap}{2}H_\r{}^{mn}R^\H_{\kb\lb mn}
+\cO(\ap^3)~.
\end{aligned}
\eeq
In the first term, \eqref{eq:W-appendix-curvature-second-order} gives $R^\H_{\kb\lb\r q}=-\half(\dd H)_{q\r\kb\lb}+\cO(\ap^2)$. In the second, it gives $R^\H_{\kb\lb mn}=\half(\dd H)_{mn\kb\lb}+\cO(\ap^2)$. Hence
\beq
\label{eq:W-geometric-two-forms}
\begin{aligned}
(H-\dd^c\o)_{\r\kb\lb}
&\=\frac{\ap}{2}\left((\nabla^-)^q+C^q-2\del^q\Phi\right)
(\dd H)_{q\r\kb\lb}\\
&\quad-\frac{\ap}{4}H_\r{}^{mn}(\dd H)_{mn\kb\lb}
+\cO(\ap^3)~.
\end{aligned}
\eeq
By the definition of $P$, the divergence and dilaton derivative give $2\ap P_{\r\kb\lb}$. This is the flux--torsion equation in \eqref{eq:W-summary-reduction}. We have kept the $C$ term: its relation to conformally balanced is given in \eqref{eq:balance-C}.

\subsubsection{The $k$ and $\ccZ^{1,1}$ stationary equations}
The $k$ variation leaves $g$, $A$ and $\cB$ fixed, so its connection recursion is homogeneous and $\d_kH=0$. For the gauge variation, $\wh\Omega-\Omega=\cO(\ap^2)$ is multiplied by the explicit $\ap$. These two coefficients therefore give
\beq
H^{0,3}\=\cO(\ap^3)~,\qquad
\ap F^{0,2}\=\cO(\ap^3)~.
\eeq
The factor of $\ap$ in the bundle equation is part of its stationary coefficient and should be retained when counting orders.

The $\ccZ^{1,1}$ equation follows from the same curvature calculation with a barred free index:
\beq
0\=2\ap P_{\rb\kb\lb}
+\frac{\ap}{2}C^m(\dd H)_{m\rb\kb\lb}
-\frac{\ap}{4}H_\rb{}^{mn}(\dd H)_{mn\kb\lb}
+\cO(\ap^3)~.
\eeq
The first two terms involve $(\dd H)^{1,3}=\cO(\ap^2)$; the connection preserves type to the required order. In the last term, the $(2,2)$ part of $\dd H$ would require $H_\rb{}^{\m\n}$, which is a component of $H^{0,3}$ after lowering indices. The remaining contractions again involve $(\dd H)^{1,3}$. Thus this coefficient gives no additional condition through $\ap^2$.

\subsubsection{The $\ccZ^{0,2}$ stationary equation}

Having dealt with the first two lines of \eqref{eq:W-summary-reduction}, we turn to the third line which is an equation for the holomorphic volume form. 

The $\ccZ^{0,2}$ and $\D$ coefficients contain the same sorts of expressions in terms of $R^\H$. Substituting the $\D$ equation into the $\ccZ^{0,2}$ coefficient, and antisymmetrising its barred indices, eliminates that expression and gives
\beq
\label{eq:W-Z02-one-form}
\delb\Omega
\=\half g^{\r\nb}(H-\dd^c\o)_{\r\nb\lb}\dd x^\lb\w\Omega~.
\eeq
This relation follows from the untruncated variation. To evaluate it through $\ap^2$, we need the Hermitian trace of \eqref{eq:W-geometric-two-forms}.

First, the lower-order volume and flux--torsion equations give
\beq
C_\mb-2\del_\mb\Phi
\=-g^{\r\nb}H_{\r\nb\mb}+\cO(\ap^2)~.
\eeq
Only this barred component enters the trace, because the leading order $\dd H$ has type $(2,2)$. Now contract $\dd(\dd H)=0$ twice with the Hermitian metric. Using $\nabla^-g=0$ and the preceding identity gives
\beq
\begin{aligned}
&g^{\r\nb}\left((\nabla^-)^m+C^m-2\del^m\Phi\right)
(\dd H)_{m\r\nb\lb}\\
&\qquad\=\half\del_\lb q_H
+\half g^{\r\nb}H_\r{}^{mn}(\dd H)_{mn\nb\lb}
+\cO(\ap^2)~.
\end{aligned}
\eeq
The $H\,\dd H$ term on the right comes from the torsion of $\nabla^-$. It cancels the explicit $H\,\dd H$ term when we trace \eqref{eq:W-geometric-two-forms}, leaving
\beq
\label{eq:trace-qH}
g^{\r\nb}(H-\dd^c\o)_{\r\nb\lb}
\=\frac{\ap}{4}\del_\lb q_H+\cO(\ap^3)~.
\eeq
Substituting into \eqref{eq:W-Z02-one-form} gives
\beq
\delb\Omega\=\frac{\ap}{8}\delb q_H\w\Omega+\cO(\ap^3)~,
\qquad
\dd\left(e^{-\ap q_H/8}\Omega\right)\=\cO(\ap^3)~.
\eeq
This is the volume-form equation in \eqref{eq:W-summary-reduction}. The exponential agrees with $1-\ap q_H/8$ to the required order because $q_H=\cO(\ap)$. Equation~\eqref{eq:W-appendix-Omega-second-order} then also identifies the rescaling with $\wh \O$ and so 
\beq
\dd\wh\Omega\=\cO(\ap^3)~.
\eeq

No D-term has been used in this trace calculation. When the lower-order D-terms are imposed, $q_H=\ap h/2+\cO(\ap^2)$ identifies the rescaling with \eqref{eq:W-volume-MP}.

\subsection{The BdR gravitino derivative convention}

The derivative $\mathcal D^-$ used by BdR acts with the minus connection on Lorentz indices and with Levi--Civita on coordinate indices. Our definition of $P$ uses $\nabla^-$ on every index. To compare the two, expand the derivative of the density and convert the remaining coordinate index to the minus connection:
\beq
\label{eq:BdR-derivative-conversion}
\begin{aligned}
&\frac{e^{2\Phi}}{\sqrt g}g_{mn}\mathcal D^-_p
\left(\sqrt g\,e^{-2\Phi}(\dd H)^{pn}{}_{ab}\right)\\
&\qquad\=\left((\nabla^-)^p-2\del^p\Phi\right)
(\dd H)_{pmab}
-\half H_m{}^{pq}(\dd H)_{pqab}\\
&\qquad\=4P_{mab}-\half H_m{}^{pq}(\dd H)_{pqab}~.
\end{aligned}
\eeq
The dilaton factor supplies $-2\del^p\Phi$. The connection change on the free coordinate index supplies the displayed $H\,\dd H$ term; the correction on the contracted divergence index vanishes by antisymmetry.

Substitution into the BdR gravitino variation gives \eqref{eq:BdR-gravitino}. Including its overall factor of $\ap$, the $H\,\dd H$ contribution is second order when $H=\cO(1)$ and third order when $H=\cO(\ap)$. It must therefore be kept in the cubic comparison even when the leading order metric is Calabi--Yau.

\subsection{Holomorphy through $\ap^3$}
\label{app:closure}
\label{app:W-holo-Z}
The goal here is to check that the antiholomorphic variation of $W$ vanishes through $\ap^3$. As a consequence,  we will also show $\dd\wh\Omega=\cO(\ap^4)$.

Consider a variation with respect to the B-field, with $g$, $A$, $\Omega$ and the vielbein fixed (so that $\d_B e^a_m=0$). Then $\d_B\Th^\H_{mpq}=\half(\d_BH)_{mpq}$, so
\beq
\label{eq:W-appendix-B-variation}
\begin{aligned}
\d_BH
&\=\dd\cB+\frac{\ap}{4}R^{\H\,pq}\w(\d_BH)_{pq}~,\\
\d_BW
&\=\int_X\dd\cB\w\wh\Omega
\=-\int_X\cB\w\dd\wh\Omega~.
\end{aligned}
\eeq
The second line follows by the same geometric series as the gauge variation. It is an all-order identity. Hence, $\d_B W = 0$  is equivalent to $\dd\wh\Omega=0$. We can't immediately conclude $\d_B W = 0$ holds  as we don't yet know that $W$ is holomorphic. We now show this is the case through $\ap^3$.

For the  $\ccZ^{0,2}$ variation of $W$, the metric stays fixed and so  $M_{pq}=\half\ccZ_{pq}$ is purely antisymmetric. With our choice of Lorentz frame $\d e_{ap}=E_a{}^qM_{qp}$ this induces a rotation of the vielbein. This rotation changes the Hull connection by minus the covariant derivative of  $\ccZ_{pq}/2$, giving
\beq\label{eq:dZTh}
\d_\ccZ\Th^\H_{mpq} \=\half(\d_\ccZ H)_{mpq}-\half\nabla^\H_m\ccZ_{pq}~.
\eeq
This is consistent with (3.6) of \cite{McOrist:2026cys} when truncated to leading order in $\ap$. We now insert \eqref{eq:dZTh} into $\d_\ccZ H=\dd\ccZ^{0,2}+\ap\,\d_\ccZ\Th^\H_{pq}\w R^{\H\,pq}/2$. The extra term from the frame rotation is
\beq
-\frac{\ap}{4}(\nabla^\H_m\ccZ_{pq})\dd x^m\w R^{\H\,pq} \=-\frac{\ap}{4}\dd\left(\ccZ_{pq}R^{\H\,pq}\right)~.
\eeq
To get the equality, we use the $R^\H$ Bianchi identity: the covariant exterior derivative of $R^\H$ vanishes, while contracting the Lorentz indices cancels the connection terms. Putting it together with $\dd\ccZ^{0,2}$ gives
\beq
\label{eq:W-Z02-recursion}
\begin{aligned}
\d_\ccZ H
&\=\dd\left(\ccZ^{0,2}
-\frac{\ap}{4}\ccZ_{\mb\nb}R^{\H\,\mb\nb}\right)
+\frac{\ap}{4}R^{\H\,pq}\w(\d_\ccZ H)_{pq}~,\\
\d_\ccZ W
&\=-\int_X\left(\ccZ^{0,2}
-\frac{\ap}{4}\ccZ_{\mb\nb}R^{\H\,\mb\nb}\right)
\w\dd\wh\Omega~.
\end{aligned}
\eeq

To conclude that $\dd\wh\Omega=0$, we need stationarity under arbitrary two-form variations. We have so far considered the holomorphic variations, so we now check the antiholomorphic ones through $\ap^3$, using the lower-order F-terms but no D-terms.

 Let $\bar\a$ label an antiholomorphic direction in field space. In holomorphic parameter gauge, $\d_{\bar\a}\Omega=0$, $\d_{\bar\a}A$ has type $(1,0)$, and $\ccZ_{\bar\a}$ has type $(2,0)$. An antiholomorphic geometric fluctuation also satisfies $M_{\mb p}=0$. Its direct contribution to \eqref{eq:W-gauge-resummed-variation_real} vanishes by type, while its curvature contribution pairs only with the $(2,3)$ part of the five-form coefficient.

To estimate that coefficient, use \eqref{eq:W-appendix-Omega-second-order}. The scalar multiple of $\Omega$ gives $R^\H\w\Omega$ of type $(3,2)$ and order $\ap$. The part of $\nabla^-$ that changes complex type is $\cO(\ap^2)$, so differentiating this term can produce a $(2,3)$ component only at $\ap^3$. The remaining correction to $\wh\Omega$ is already $\cO(\ap^3)$. Derivatives on the tangent indices and the explicit $H$ term do not change form type. Hence
\beq
\label{eq:W-closure-opposite-type-order}
\left[
\nabla^-_q\left(R^{\H\,pq}\w\wh\Omega\right)
+\half H_{rs}{}^pR^{\H\,rs}\w\wh\Omega
\right]^{2,3}_J
\=\cO(\ap^3)~.
\eeq
Its explicit factor of $\ap$ in the variation gives $\d_{\bar\a}W=\cO(\ap^4)$ in the geometric directions. For the bundle directions, the scalar multiple of $\Omega$ again gives zero by type, while the remaining $\cO(\ap^3)$ correction is multiplied by $\ap$. Thus
\beq
\label{eq:W-appendix-holomorphy-cubic}
\d_{\bar\a}W\=\cO(\ap^4)~.
\eeq
Hence, \eqref{eq:W-appendix-B-variation}  implies
\beq
\dd\wh\Omega\=\cO(\ap^4)~.
\eeq
This supplies the closedness used in \eqref{eq:W-cubic-volume}. 

There remain calculations to check. So far, we have checked just one derivative; we really need holomorphy in an neighbourhood. We also plan to extend this to a manifestly all order calculation. However, the fact it works up to and including $\ap^3$ combined with the holomorphic stationary equations reproducing BdR \citeBdR is compelling evidence that $W$ is indeed the correct holomorphic superpotential for this setting.

\section{A comparison with four-dimensional supersymmetry}\label{app:four-dimensional}

In rigid $d \= 4$, $\mathcal N \= 1$ supersymmetry, the fermion variations on a bosonic configuration with constant scalars and vanishing gauge field strengths take the form
\beq
\delta\chi^i \= \epsilon F^i,\qquad \delta\lambda^a \= \frac{1}{\sqrt{2}}\epsilon D^a,\qquad F^i \= -K^{i\bar j}\partial_{\bar j}\overline W~.
\eeq
In this rigid theory, the scalar potential is a sum of squares. A supersymmetric vacuum has $F^i = 0$ and $D^a = 0$.

A gauge-covariant, possibly field-dependent fermion redefinition gives
\beq
\widetilde\chi^i \= \chi^i + \sqrt{2}\,M^i{}_a\lambda^a~, \qquad \delta\widetilde\chi^i \= \epsilon\bigl(F^i + M^i{}_aD^a\bigr)~,
\eeq
since $(\delta M^i{}_a)\lambda^a$ vanishes on a bosonic background. Keeping the gaugino unchanged makes this redefinition invertible, so
\beq
\{F + MD \= 0,\ D \= 0\} \= \{F \= 0,\ D \= 0\}~.
\eeq
Thus a selected fermion equation can acquire a D-term contribution while the simultaneous supersymmetry conditions remain unchanged.

This changes the fermion basis. It does not change the equations $\partial_i W = 0$ or the simultaneous zero of the F-terms and D-terms. It explains why the D-term on the right of \eqref{eq:W-summary-reduction_final} is consistent with the same supersymmetric vacua.

\section{Details on $\ap^3$ calculation}

\subsection{Flux torsion calculation}
\label{s:fluxtorsionap3}
We give the intermediate steps from \eqref{eq:W-Delta-cubic-start} to \eqref{eq:W-Delta-MP-cubic}. Throughout, we use $H,\dd\Phi,P=\cO(\ap)$. The lower-order flux--torsion equation \eqref{eq:W-summary-torsion} gives the connection correction in \eqref{eq:minusconnection},
\beq
\Gamma^- \= \Gamma^{\mathrm B}-\ap P+\cO(\ap^3)~.
\eeq
Since the Bismut connection preserves complex type, $R^{\mathrm B}_{pq\kb\lb}=0$. Expanding the minus curvature therefore gives
\beq
\label{eq:W-appendix-minus-curvature}
R^-_{pq\kb\lb}
\= -\ap\left(\nabla^-_pP_{q\kb\lb}-\nabla^-_qP_{p\kb\lb}\right)+\cO(\ap^3)~.
\eeq
The torsion term in this expansion is proportional to $\ap HP=\cO(\ap^3)$, while the term quadratic in the connection correction is $\ap^2P^2=\cO(\ap^4)$. Neither contributes to the stationary equation through $\ap^3$. The pair exchange identity \eqref{eq:pair} now gives
\beq
\label{eq:W-appendix-Hull-curvature}
\begin{aligned}
R^{\mathrm H}_{\kb\lb\r q}
&\=R^-_{\r q\kb\lb}+\half(\dd H)_{\kb\lb\r q}\\
&\=-\half(\dd H)_{q\r\kb\lb}
-\ap\left(\nabla^-_\r P_{q\kb\lb}-\nabla^-_qP_{\r\kb\lb}\right)
+\cO(\ap^3)~.
\end{aligned}
\eeq
The sign in the first term comes from moving $q$ to the first slot of $\dd H$.

In the curvature term of \eqref{eq:W-Delta-cubic-start}, we may replace $\wh\Omega$ by $\Omega$ at the expense of $\ap^4$ errors. Using
\beq
\nabla^{\mathrm B}_q\Omega
\= -2(\partial_q\Phi)\Omega+\cO(\ap^2)~,
\eeq
the component form of \eqref{eq:W-Delta-cubic-start} is
\beq
\label{eq:W-appendix-Delta-components}
\begin{aligned}
(H-\dd^c\o)_{\r\kb\lb}
&\=-\ap e^{2\Phi}(\nabla^{\mathrm B})^q
\left(e^{-2\Phi}R^{\mathrm H}_{\kb\lb\r q}\right)\\
&\quad-\frac{\ap}{2}H_\r{}^{mn}R^{\mathrm H}_{\kb\lb mn}
+\cO(\ap^4)~.
\end{aligned}
\eeq

Substituting \eqref{eq:W-appendix-Hull-curvature} gives
\beq
\label{eq:W-appendix-Delta-expanded}
\begin{aligned}
(H-\dd^c\o)_{\r\kb\lb}
&\=\frac{\ap}{2}e^{2\Phi}(\nabla^-)^q
\left(e^{-2\Phi}\dd H\right)_{q\r\kb\lb}
-\frac{\ap}{4}H_\r{}^{mn}(\dd H)_{mn\kb\lb}\\
&\quad+\ap^2(\nabla^-)^q
\left(\nabla^-_\r P_{q\kb\lb}-\nabla^-_qP_{\r\kb\lb}\right)
+\cO(\ap^4)~.
\end{aligned}
\eeq
Here replacing $\nabla^{\mathrm B}$ by $\nabla^-$ changes the $\dd H$ term only at $\cO(\ap^4)$. The dilaton derivative multiplying the curl of $P$ is also $\cO(\ap^4)$, as is the contribution of that curl to the term containing $H_\r{}^{mn}$.

By the definition of $P$,
\beq
P_{\r\kb\lb}
\=\frac14e^{2\Phi}(\nabla^-)^q
\left(e^{-2\Phi}\dd H\right)_{q\r\kb\lb}~.
\eeq
Thus the first term on the right hand side of \eqref{eq:W-appendix-Delta-expanded} is $2\ap P_{\r\kb\lb}$. Moving it and the $H\,\dd H$ term to the left gives
\beq
\begin{aligned}
&(H-\dd^c\o-2\ap P)_{\r\kb\lb}
+\frac{\ap}{4}H_\r{}^{mn}(\dd H)_{mn\kb\lb}\\
&\qquad\=\ap^2(\nabla^-)^q
\left(\nabla^-_\r P_{q\kb\lb}-\nabla^-_qP_{\r\kb\lb}\right)
+\cO(\ap^4)~,
\end{aligned}
\eeq
which is \eqref{eq:W-Delta-MP-cubic}.

\subsection{The volume-form expansion}
\label{s:vol_form}
We spell out how the geometric series gives \eqref{eq:W-cubic-volume-expansion}. We use the same lower-order F-terms and D-terms as in the main text, so $H,P=\cO(\ap)$ and the leading order metric is Calabi--Yau. Our starting point is
\beq
\wh\Omega \= \Omega+\frac{\ap}{4}\cK\Omega
+\frac{\ap^2}{16}\cK^2\Omega+\cdots~,
\qquad
\cK\Omega \= (R^{\H\,pq}\w\Omega)_{pq}~.
\eeq
We first calculate $\cK\Omega$ through $\ap^2$, then check which further terms in the series can contribute.

Since $\Omega$ has type $(3,0)$, only the $(0,2)$ part of the curvature survives in $R^{\H\,pq}\w\Omega$. The Bismut connection preserves complex type, so $R^{\mathrm B}_{pq\kb\lb}=0$. Pair exchange \eqref{eq:pair} and the minus connection defined in  \eqref{eq:minusconnection} therefore give
\beq\label{eq:RHdHNablaP}
R^\H_{\kb\lb pq}
\= \half(\dd H)_{\kb\lb pq}
-\ap\left(\nabla_pP_{q\kb\lb}-\nabla_qP_{p\kb\lb}\right)
+\cO(\ap^3)~.
\eeq
The derivatives here can use the leading order Calabi--Yau connection, since $\ap P=\cO(\ap^2)$. The lower-order flux--torsion equation and the $\delb$-exact leading order part of $P^{1,2}$ in \eqref{eq:W-Delta-P-cubic} give $(\dd H)^{1,3}=\cO(\ap^3)$. Since the part of $\nabla^-$ that changes complex type starts at $\ap^2$, the definition of $P$ then also gives $P^{0,3}=\cO(\ap^3)$.

Taking the two contractions in $\cK$, we find
\beq
\cK\Omega
\= R^\H_{\mb\nb}{}^{\mb\nb}\Omega
+2R^\H_{\nb\lb}{}^{\r\nb}\dd x^\lb\w\Omega_\r
+\cO(\ap^3)~,
\eeq
where \eqref{eq:RHdHNablaP} gives
\beq
\begin{aligned}
R^\H_{\mb\nb}{}^{\mb\nb}
&\= -\half q_H
+2\ap\nabla^\lb\left(g^{\m\nb}P_{\m\nb\lb}\right)
+\cO(\ap^3)~,\\
R^\H_{\nb\lb}{}^{\r\nb}
&\= -\ap\nabla^\r\left(g^{\m\nb}P_{\m\nb\lb}\right)
+\cO(\ap^3)~.
\end{aligned}
\eeq
The scalar contraction of $\dd H$ gives $-q_H/2$, while the two derivatives of $P$ give equal contributions to its trace. A possible $(1,2)$ term would involve $R^\H_{\kb\lb}{}^{\m\n}$, which is already $\cO(\ap^3)$.

This also tells us where to stop the geometric series. We have $\cK\Omega=-q_H\Omega/2+\cO(\ap^2)$, with $q_H=\cO(\ap)$. Since $\cK$ is algebraic and does not differentiate $q_H$, a second application gives $\cK^2\Omega=-q_H\cK\Omega/2+\cO(\ap^2)=\cO(\ap^2)$. Its prefactor therefore makes it fourth order, and further powers cannot contribute through $\ap^3$.

Now insert the two identities \eqref{eq:W-cubic-volume-contractions},
\beq
q_H \= \frac{\ap}{2}h+\cO(\ap^3)~,
\qquad
g^{\m\nb}P_{\m\nb\lb}
\= \frac{\ap}{16}\del_\lb h+\cO(\ap^2)~.
\eeq
Using $\nabla^\lb\del_\lb h=\half\nabla^q\nabla_qh$ and commuting the barred derivatives of $h$ gives
\beq
\begin{aligned}
\wh\Omega
&\= \left(1-\frac{\ap^2h}{16}
+\frac{\ap^3}{64}\nabla^q\nabla_qh\right)\Omega\\
&\quad-\frac{\ap^3}{32}\delb(\nabla^\r h)\w\Omega_\r
+\cO(\ap^4)~.
\end{aligned}
\eeq
All derivatives and contractions in the explicitly cubic terms use the leading order Calabi--Yau metric. At this order $\Omega$ is parallel, so
\beq
\del\left[(\nabla^\r h)\Omega_\r\right]
\= (\nabla_\r\nabla^\r h)\Omega
\= \half(\nabla^q\nabla_qh)\Omega~.
\eeq
Writing $\delb=\dd-\del$ on this two-form gives \eqref{eq:W-cubic-volume-expansion}: the scalar Laplacian coefficient becomes $\ap^3/32$, and the remaining correction is an exact three-form.


\begin{thebibliography}{10}

\bibitem{Bergshoeff:1988nn}
E.~Bergshoeff and M.~de~Roo, {\it {Supersymmetric Chern-Simons Terms in
  Ten-Dimensions}},  {\em Phys.Lett.} {\bf B218} (1989) 210.

\bibitem{Bergshoeff:1989de}
E.~Bergshoeff and M.~de~Roo, {\it {The Quartic Effective Action of the
  Heterotic String and Supersymmetry}},  {\em Nucl.Phys.} {\bf B328} (1989)
  439.

\bibitem{Hull:1986kz}
C.~Hull, {\it {Compactifications of the Heterotic Superstring}},  {\em
  Phys.Lett.} {\bf B178} (1986) 357.

\bibitem{Strominger:1986uh}
A.~Strominger, {\it {Superstrings with Torsion}},  {\em Nucl. Phys. B} {\bf
  274} (1986) 253.

\bibitem{Becker:2003gq}
K.~Becker, M.~Becker, K.~Dasgupta, and S.~Prokushkin, {\it {Properties of
  heterotic vacua from superpotentials}},  {\em Nucl. Phys. B} {\bf 666} (2003)
  144--174, [\href{http://xxx.lanl.gov/abs/hep-th/0304001}{{\tt
  hep-th/0304001}}].

\bibitem{LopesCardoso:2003dvb}
G.~Lopes~Cardoso, G.~Curio, G.~Dall'Agata, and D.~Lust, {\it {BPS action and
  superpotential for heterotic string compactifications with fluxes}},  {\em
  JHEP} {\bf 10} (2003) 004,
  [\href{http://xxx.lanl.gov/abs/hep-th/0306088}{{\tt hep-th/0306088}}].

\bibitem{Gurrieri:2004dt}
S.~Gurrieri, A.~Lukas, and A.~Micu, {\it {Heterotic on half-flat}},  {\em
  Phys.Rev.} {\bf D70} (2004) 126009,
  [\href{http://xxx.lanl.gov/abs/hep-th/0408121}{{\tt hep-th/0408121}}].

\bibitem{deRoo:1992zp}
M.~de~Roo, H.~Suelmann, and A.~Wiedemann, {\it {The Supersymmetric effective
  action of the heterotic string in ten-dimensions}},  {\em Nucl. Phys. B} {\bf
  405} (1993) 326--366, [\href{http://xxx.lanl.gov/abs/hep-th/9210099}{{\tt
  hep-th/9210099}}].

\bibitem{McOrist:2025zwf}
J.~McOrist and S.~Picard, {\it {Stringy Corrections to Heterotic
  SU(3)-Geometry}},  {\em Commun. Math. Phys.} {\bf 407} (2026), no.~5 105,
  [\href{http://xxx.lanl.gov/abs/2507.0238}{{\tt arXiv:2507.0238}}].

\bibitem{MPap3}
J.~McOrist and S.~Picard, {\it On $\alpha'^3$ corrections in heterotic string
  theory},  {\em work in progress} (2026).

\bibitem{Coimbra:2014geometry}
A.~Coimbra, R.~Minasian, H.~Triendl, and D.~Waldram, {\it Generalised geometry
  for string corrections},  {\em JHEP} {\bf 11} (2014) 160,
  [\href{http://xxx.lanl.gov/abs/1407.7542}{{\tt arXiv:1407.7542}}].

\bibitem{Smith:2024moduli}
G.~R. Smith, D.~Tennyson, and D.~Waldram, {\it All-orders moduli for type ii
  flux backgrounds},  \href{http://xxx.lanl.gov/abs/2409.0384}{{\tt
  arXiv:2409.0384}}.

\bibitem{liu2025typeiibderivativesfivepoint}
J.~Liu, R.~Minasian, R.~Savelli, and A.~Schachner, {\it Type iib at eight
  derivatives: Five-point axio-dilaton couplings},  2025.

\bibitem{Ashmore:2026vgf}
A.~Ashmore and S.~Sethi, {\it {A real superpotential for M-theory on
  eight-manifolds}},  \href{http://xxx.lanl.gov/abs/2610.1243}{{\tt
  arXiv:2610.1243}}.

\bibitem{Gukov:1999ya}
S.~Gukov, C.~Vafa, and E.~Witten, {\it {CFT's from Calabi-Yau four folds}},
  {\em Nucl. Phys. B} {\bf 584} (2000) 69--108,
  [\href{http://xxx.lanl.gov/abs/hep-th/9906070}{{\tt hep-th/9906070}}].
  [Erratum: Nucl.Phys.B 608, 477--478 (2001)].

\bibitem{Becker:2002jj}
M.~Becker and D.~Constantin, {\it {A Note on flux induced superpotentials in
  string theory}},  {\em JHEP} {\bf 08} (2003) 015,
  [\href{http://xxx.lanl.gov/abs/hep-th/0210131}{{\tt hep-th/0210131}}].

\bibitem{Becker:2003yv}
K.~Becker, M.~Becker, K.~Dasgupta, and P.~S. Green, {\it {Compactifications of
  heterotic theory on nonKahler complex manifolds. 1.}},  {\em JHEP} {\bf 04}
  (2003) 007, [\href{http://xxx.lanl.gov/abs/hep-th/0301161}{{\tt
  hep-th/0301161}}].

\bibitem{McOrist:2016cfl}
J.~McOrist, {\it {On the Effective Field Theory of Heterotic Vacua}},  {\em
  Lett. Math. Phys.} {\bf 108} (2018), no.~4 1031--1081,
  [\href{http://xxx.lanl.gov/abs/1606.0522}{{\tt arXiv:1606.0522}}].

\bibitem{Sethi:2017phn}
S.~Sethi, {\it {Supersymmetry Breaking by Fluxes}},  {\em JHEP} {\bf 10} (2018)
  022, [\href{http://xxx.lanl.gov/abs/1709.0355}{{\tt arXiv:1709.0355}}].

\bibitem{delaOssa:2015maa}
X.~de~la Ossa, E.~Hardy, and E.~E. Svanes, {\it {The Heterotic Superpotential
  and Moduli}},  {\em JHEP} {\bf 01} (2016) 049,
  [\href{http://xxx.lanl.gov/abs/1509.0872}{{\tt arXiv:1509.0872}}].

\bibitem{Ashmore:2018ybe}
A.~Ashmore, X.~De~La~Ossa, R.~Minasian, C.~Strickland-Constable, and E.~E.
  Svanes, {\it {Finite deformations from a heterotic superpotential:
  holomorphic Chern-Simons and an $L_\infty$ algebra}},  {\em JHEP} {\bf 10}
  (2018) 179, [\href{http://xxx.lanl.gov/abs/1806.0836}{{\tt
  arXiv:1806.0836}}].

\bibitem{Ashmore:2019rkx}
A.~Ashmore, C.~Strickland-Constable, D.~Tennyson, and D.~Waldram, {\it
  {Heterotic backgrounds via generalised geometry: moment maps and moduli}},
  {\em JHEP} {\bf 11} (2020) 071,
  [\href{http://xxx.lanl.gov/abs/1912.0998}{{\tt arXiv:1912.0998}}].

\bibitem{Kupka:2024rvl}
J.~Kupka, C.~Strickland-Constable, E.~E. Svanes, D.~Tennyson, and F.~Valach,
  {\it {BPS complexes and Chern--Simons theories from $G$-structures in gauge
  theory and gravity}},  \href{http://xxx.lanl.gov/abs/2406.0355}{{\tt
  arXiv:2406.0355}}.

\bibitem{Ashmore:2023vji}
A.~Ashmore, J.~J.~M. Ibarra, D.~D. McNutt, C.~Strickland-Constable, E.~E.
  Svanes, D.~Tennyson, and S.~Winje, {\it {A heterotic Kodaira-Spencer theory
  at one-loop}},  {\em JHEP} {\bf 10} (2023) 130,
  [\href{http://xxx.lanl.gov/abs/2306.1010}{{\tt arXiv:2306.1010}}].

\bibitem{delaOssa:2014cia}
X.~de~la Ossa and E.~E. Svanes, {\it {Holomorphic Bundles and the Moduli Space
  of N=1 Supersymmetric Heterotic Compactifications}},  {\em JHEP} {\bf 10}
  (2014) 123, [\href{http://xxx.lanl.gov/abs/1402.1725}{{\tt
  arXiv:1402.1725}}].

\bibitem{McOrist:2021dnd}
J.~McOrist and E.~E. Svanes, {\it {Heterotic quantum cohomology}},  {\em JHEP}
  {\bf 11} (2022) 096, [\href{http://xxx.lanl.gov/abs/2110.0654}{{\tt
  arXiv:2110.0654}}].

\bibitem{Anguelova:2010ed}
L.~Anguelova, C.~Quigley, and S.~Sethi, {\it {The Leading Quantum Corrections
  to Stringy Kahler Potentials}},  {\em JHEP} {\bf 1010} (2010) 065,
  [\href{http://xxx.lanl.gov/abs/1007.4793}{{\tt arXiv:1007.4793}}].

\bibitem{Candelas:2016usb}
P.~Candelas, X.~de~la Ossa, and J.~McOrist, {\it {A Metric for Heterotic
  Moduli}},  {\em Commun. Math. Phys.} {\bf 356} (2017), no.~2 567--612,
  [\href{http://xxx.lanl.gov/abs/1605.0525}{{\tt arXiv:1605.0525}}].

\bibitem{McOrist:2019mxh}
J.~McOrist and R.~Sisca, {\it {Small gauge transformations and universal
  geometry in heterotic theories}},  {\em SIGMA} {\bf 16} (2020) 126,
  [\href{http://xxx.lanl.gov/abs/1904.0757}{{\tt arXiv:1904.0757}}].

\bibitem{McOrist:2024zdz}
J.~McOrist, S.~Picard, and E.~E. Svanes, {\it {A Heterotic Hermitian-Yang-Mills
  Equivalence}},  {\em Commun. Math. Phys.} {\bf 406} (2025), no.~5 107,
  [\href{http://xxx.lanl.gov/abs/2402.1035}{{\tt arXiv:2402.1035}}].

\bibitem{Melnikov:2014ywa}
I.~V. Melnikov, R.~Minasian, and S.~Sethi, {\it {Heterotic fluxes and
  supersymmetry}},  {\em JHEP} {\bf 06} (2014) 174,
  [\href{http://xxx.lanl.gov/abs/1403.4298}{{\tt arXiv:1403.4298}}].

\bibitem{Becker:2009df}
K.~Becker and S.~Sethi, {\it {Torsional Heterotic Geometries}},  {\em Nucl.
  Phys.} {\bf B820} (2009) 1--31,
  [\href{http://xxx.lanl.gov/abs/0903.3769}{{\tt arXiv:0903.3769}}].

\bibitem{Chapline:1982ww}
G.~Chapline and N.~Manton, {\it {Unification of Yang-Mills Theory and
  Supergravity in Ten-Dimensions}},  {\em Phys.Lett.} {\bf B120} (1983)
  105--109.

\bibitem{Hitchin:2000jd}
N.~J. Hitchin, {\it {The Geometry of Three-Forms in Six Dimensions}},  {\em J.
  Diff. Geom.} {\bf 55} (2000), no.~3 547--576,
  [\href{http://xxx.lanl.gov/abs/math/0010054}{{\tt math/0010054}}].

\bibitem{Candelas:2018lib}
P.~Candelas, X.~De~La~Ossa, J.~McOrist, and R.~Sisca, {\it {The Universal
  Geometry of Heterotic Vacua}},  {\em JHEP} {\bf 02} (2019) 038,
  [\href{http://xxx.lanl.gov/abs/1810.0087}{{\tt arXiv:1810.0087}}].

\bibitem{McOrist:2026boe}
J.~McOrist, {\it {Universal geometry as an organising principle for heterotic
  moduli}},  \href{http://xxx.lanl.gov/abs/2606.3119}{{\tt arXiv:2606.3119}}.

\bibitem{McOrist:2026cys}
J.~McOrist and Q.~Yin, {\it {Heterotic moduli, the double extension and the
  alpha'{\textasciicircum}2 metric}},
  \href{http://xxx.lanl.gov/abs/2607.1581}{{\tt arXiv:2607.1581}}.

\end{thebibliography}

\providecommand{\href}[2]{#2}\begingroup\raggedright\endgroup

\end{document}